 \documentclass{aastex}

\newcommand \lta {\mathrel{\vcenter
     {\hbox{$<$}\nointerlineskip\hbox{$\sim$}}}}
\newcommand \gta {\mathrel{\vcenter
     {\hbox{$>$}\nointerlineskip\hbox{$\sim$}}}}

\newcommand\kms{km~s$^{-1}$}

\slugcomment{draft}

\begin{document}

\title{The Magnetorotational Instability in Core Collapse Supernova 
Explosions}

\author{Shizuka Akiyama, J. Craig Wheeler}
\affil{Astronomy Department, University of Texas, Austin, TX 78712}
\email{shizuka@astro.as.utexas.edu,wheel@astro.as.utexas.edu}

\author{David L. Meier}
\affil{Jet Propulsion Laboratory, 4800 Oak Grove Dr., Pasadena, CA 91109}
\email{dlm@sgra.jpl.nasa.gov}

\and

\author{Itamar Lichtenstadt}
\affil{Hebrew University of Jerusalem}
\email{itamar@wiener.fiz.huji.ac.il}

\begin{abstract}

We investigate the action of the magnetorotational instability (MRI) in
the context of iron--core collapse.  Exponential growth of the field on
the time scale $\Omega^{-1}$ by the MRI will dominate the linear growth
process of field line ``wrapping" with the same characteristic time.  We
examine a variety of initial rotation states, with solid body rotation or
a gradient in rotational velocity, that correspond to models in the
literature.  A relatively modest value of the initial rotation, a period
of $\sim$ 10 s, will give a very rapidly rotating PNS and hence strong
differential rotation with respect to the infalling matter. We assume
conservation of angular momentum on spherical shells.  Rotational
distortion and the dynamic feedback of the magnetic field are neglected in
the subsequent calculation of rotational velocities.  In our rotating and
collapsing conditions, a seed field is expected to be amplified by the MRI
and to grow exponentially to a saturation field.  Results are discussed
for two examples of saturation fields, a fiducial field that corresponds
to $v_{\rm{A}} = r\Omega$ and a field that corresponds to the maximum
growing mode of the MRI.  We find, as expected, that the shear is strong
at the boundary of the newly formed protoneutron star, and, unexpectedly,
that the region within the stalled shock can be subject to strong MHD
activity.  Modest initial rotation velocities of the iron core result in
sub--Keplerian rotation and a sub--equipartition magnetic field that
nevertheless produce substantial MHD luminosity and hoop stresses :
saturation fields of order $10^{14}$ -- $10^{16}$ G can develop $\sim$ 300
msec after bounce with an associated MHD luminosity of $\sim 10^{52}$ erg
s$^{-1}$.  Bi-polar flows driven by this MHD power can affect or even
cause the explosions associated with core-collapse supernovae.

\end{abstract}


\section{Introduction}

There is accumulating evidence that core collapse supernovae are
distinctly and significantly asymmetric.  A number of supernova remnants
show intrinsic ``bilateral" structure (Dubner et al. 2002). Jet and
counter jet structures have been mapped for Cas A in the optical (Fesen \&
Gunderson 1996; Fesen 2001; and references therein).  Chandra X-ray
Observatory (CXO) and XMM Newton observations of Cas A show that the jet
and counter jet and associated structure are observable in the X-ray and
that the intermediate mass elements are ejected in a roughly toroidal
configuration (Hughes et al. 2000;  Hwang et al. 2000; Willingale et al.
2002).  HST observations of the debris of SN~1987A show that the ejecta
are asymmetric with an axis that roughly aligns with the small axis of the
rings (Pun et al. 2001; Wang et al. 2002).  
Spectropolarimetry shows that substantial
asymmetry is ubiquitous in core-collapse supernovae, and that a
significant fraction of core-collapse supernovae have a bi-polar structure
(Wang et al. 1996, 2001).  The strength of the asymmetry observed with
polarimetry is higher (several \%)  in supernovae of Type Ib and Ic that
represent exploding bare non-degenerate cores (Wang et al. 2001).  The
degree of asymmetry also rises as a function of time for Type II
supernovae (from $\lta$ 1\% to $\gta$ 1\%) as the ejecta expand and the
photosphere recedes (Wang et al.  2001; Leonard et al. 2000, 2001).  Both
of these trends suggest that it is the core collapse mechanism itself that
is responsible for the asymmetry.

Two possibilities are being actively explored to account for the observed
asymmetries.  One is associated with the rotational effect on convection
(Fryer \& Heger 2000) and another is due to the effect of jets (Khokhlov
et al. 1999; Wheeler et al. 2000; Wheeler, Meier \& Wilson 2002).  The
time scale of neutrino emission is short compared to the dynamical time of
the overlying stellar mantle and envelope. Neutrino asymmetries can yield
short-lived, impulsive effects (Shimizu, et al. 1994; Burrows \& Hayes
1996; Lai et al. 2001), but there are questions of whether expansion and
transverse pressure gradients will eliminate transient asymmetries before
homologous expansion is achieved (Chevalier \& Soker 1989).  
Rayleigh--Taylor and Richtmyer--Meshkov instabilities might produce
``finger'' asymmetries that are preserved, but it is unclear that such
finer scale perturbations can reproduce the common feature of a single
symmetry axis that is substantially independent of space and time (Wang et
al. 2001). Jet calculations have established that non-relativistic axial
jets of energy of order $10^{51}$ erg originating within the collapsed
core can initiate a bi-polar asymmetric supernova explosion that is
consistent with the spectropolarimetry (Khokhlov et al. 1999; Khokhlov \&
H\"oflich 2001; H\"oflich et al. 2001). The result is that heavy elements
(e.g. O, Ca) are characteristically ejected in tori along the equator.
Iron, silicon and other heavy elements in Cas A are distributed in this
way (Hwang et al. 2000), and there is some evidence for this distribution
in SN 1987A (Wang et al. 2002).  Radioactive matter ejected in the jets
can alter the ionization structure and hence the shape of the photosphere
of the envelope even if the density structure is spherically symmetric
(H\"oflich et al. 2001).  This will generate a finite polarization, even
though the density distribution is spherical and the jets are stopped deep
within the star and may account for the early polarization observed in
Type II supernovae (Leonard et al. 2000; Wang et al. 2001).  If one of the
pair of axial jets is somewhat stronger than the other, jets can, in
principle, also account for pulsar runaway velocities that are parallel to
the spin axis (Helfand et al. 2001, and references therein).  While a
combination neutrino--induced/jet--induced explosion may prove necessary
for complete understanding of core-collapse explosions, jets of the
strength computed by Khokhlov et al. (1999) are sufficient for supernova
explosions.  In this paper, we will explore the possible conditions that
could lead to the formation of buoyant bi-polar MHD outflow.

Immediately after the discovery of pulsars there were suggestions that
rotation and magnetic fields could be a significant factor in the
explosion mechanism (Ostriker \& Gunn 1971; Bisnovatyi-Kogan 1971;  
Bisnovatyi-Kogan \& Ruzmaikin 1976; Kundt 1976).  Typical dipole fields of
$10^{12}$ G and rotation periods of several to several tens of
milliseconds yield electrodynamic power of $\sim 10^{44-45}$ erg s$^{-1}$
that is insufficient to produce a strong explosion.  The evidence for
asymmetries and the possibility that bi-polar flows or jets can 
account for the observations
suggest that this issue must be revisited.  The fact that pulsars like
those in the Crab and Vela remnants have jet-like protrusions (Weisskopf
et al. 2000; Helfand et al. 2001) also encourages this line of thought.  
The present-day jets in young pulsars may be vestiges of much more
powerful MHD jets that occurred when the pulsar was born.  The transient
values of the magnetic field and rotation could have greatly exceeded
those observed today.  Tapping that energy to power the explosion could be
the very mechanism that results in the modest values of rotation
and field the pulsars display after the ejecta have dispersed.


In the current context, one would like to know not only whether or not the
rotation and magnetic field of a nascent neutron star can power the
supernova explosion, but, specifically, whether or not bi-polar flows or
jets form and whether or not they are sufficiently energetic to drive the
explosion. Possible physical mechanisms for inducing axial flows,
asymmetric supernovae, and related phenomena driven by magnetorotational
effects were considered by Wheeler et al. (2000).  The means of amplifying
magnetic fields by linear wrapping associated with differential rotation
in the neutron star and possibly by an $\alpha$~--~$\Omega$ dynamo were
discussed. Whereas Wheeler et al. (2000) focused on the effect of the
resulting net dipole field, Wheeler, Meier, \& Wilson (2002) recognized
that the toroidal field would be the dominant component of the magnetic
field and explored the capacity of the toroidal field to directly generate
axial jets by analogy with magneto-centrifugal models of jets in active
galactic nuclei (Koide et al. 2000; and references therein). Wheeler,
Meier, \& Wilson (2002) found that the production of a strong toroidal
field, substantially stronger than the $10^{12}$ G field of a pulsar, and
strong axial jets driven by that field are possible.

A caveat to consider is that Wheeler, Meier, \& Wilson (2002) again
considered only amplification of the field by ``wrapping," a process that
only increases the field linearly, and hence rather slowly in time.  
While conditions of very rapid rotation might exist that lead to a
sufficiently rapid growth of the magnetic field, that is not guaranteed.  
In addition, reconnection might limit the field before it can be
wrapped the thousands of times necessary to be interesting.
Here we consider the effects of magnetic shearing, the magnetorotational
instability (MRI; Balbus \& Hawley 1991, 1998), on the strongly shearing
environment that must exist in a nascent neutron star.  This instability
is expected to lead to the rapid exponential growth of the magnetic field with
characteristic time scale of order the rotational period.  While this
instability has been widely explored in the context of accretion disks,
this is the first time it has been applied to core collapse.  We will
argue that this instability must inevitably occur in core collapse, that
it is likely to be the dominant mechanism for the production of magnetic
flux in the context of core collapse, and that it has the capacity
to produce fields that are sufficiently strong to affect, if not cause, 
the explosion.

In \S 2 we describe our assumptions by which we treat the angular momentum
and field amplification in the context of a numerical collapse
calculation.  In \S 3, we present the results of these calculations.
Section 4 presents a discussion and conclusions.

\section{Method}

The collapse of a model iron core of a 15 M$_{\sun}$ progenitor is
simulated with a one-dimensional flux-limited diffusion code
\citep{itamar87} that computes radial profiles for various quantities.  
The core collapse code does not include rotation, convection, or magnetic
fields.  To treat rotation of the iron core, we attribute various initial
rotational profiles to the initial density profile.  The evolution of the
angular velocity profile $\Omega(r)$ is then computed using the radial
density profiles produced by the core collapse code assuming that the
specific angular momentum of a given shell is constant (\S 2.1).  The
magnetic field is obtained using the $\rho(r)$, $\Omega(r)$, and
$d\Omega/dr$ profiles according to the theory of the MRI (\S 2.2). The
dynamical feedback of the magnetic field through mechanisms such as
magnetic pressure, torque, and hoop stress are neglected here.  The MHD
luminosity is estimated using the magnetic field calculated (\S 2.3).

In this calculation, the initial iron core of 1.5 M$_{\sun}$ forms a
collapsing homologous core of mass 0.69 M$_{\sun}$.  The bounce occurs
about 166 ms after the onset of collapse and the shock forms at the
boundary of the initial homologous core at a radius of 13.8 km.  After
bounce, the outer boundary of the protoneutron star (PNS) extends to about
80 -- 90 km.  The bounce shock reaches 500 km at about 120 ms after bounce
where it stalls and remains at about this radius for the duration of the
calculation.  Between the boundary of the PNS and the shock radius, there
is a region of transition where shock--heated material settles onto the
PNS.  By 210 ms after bounce, the PNS develops a core in hydrostatic
equilibrium (v $\ll$ 100 \kms) with radius 9 km and mass 0.2 M$_\sun$.  
The hydrostatic core grows out to 24.4 km and 0.69 M$_\sun$ by 350 ms and
stays constant to the end of the simulation.  The hydrostatic core mass of
the PNS at this time encompasses all the initial homologous core mass, 0.69
M$_\sun$.  Although the original homologous core and the hydrostatic core
have the same mass, the size of the PNS core is larger because the density
decreases after bounce.

\subsection{Angular Velocity Profile}

The evolution of massive stars can be significantly affected by their
rotation \citep{eddi25,kip70,endal76,endal78,kip89,mey00,mae00},
metallicity \citep{meynet94}, and mass loss \citep{chiosi86}.  Massive
stars that are the progenitors of core collapse supernovae are known to be
rapid rotators as main sequence stars with mean $v\sin{i}$ ranging from
100 km s$^{-1}$ to 300 km s$^{-1}$
\citep{slet56,conti77,fukuda82,penny96,howarth97}.  \citet{mae99} showed
that the number ratio of Be to B-type stars increases as the local
metallicity decreases, suggesting that massive stars rotate faster in
lower metallicity environments.  According to the parameterization of mass
loss rates employed by \citet{HLW00}, the mass loss rates are enhanced by
faster rotation and higher metallicity.  Evolved stars are also observed
to have higher mass loss rates.  These effects are interdependent.  
Although mass loss significantly influences the evolutionary tracks in the
HR diagram, it does not affect the evolution of the central core since
after the He burning phase the core evolves independently from the
envelope \citep{chiosi86}.  Effects of the magnetic field on stellar
evolution \citep{HW02} may influence the timing of the decoupling
between the core and the envelope \citep{FH00}, which
could affect the final angular momentum of the core.  It is
important to understand the evolution in both the envelope and the core of
the progenitor star up to the precollapse stage since the outcome of core
collapse may depend significantly on the precollapse conditions.  For our
study, the structure of the precollapse iron core, including the angular
momentum distribution, is of particular significance.

The first calculations of the redistribution of angular momentum in the
evolution of rotating massive stars were carried by \citet{endal78}, who
obtained angular velocity profiles up to the ignition of carbon burning.  
The evolution of the angular momentum distribution of massive stars up to
the onset of core collapse was recently simulated by \citet{HLW00} for
supernova progenitors (8$M_{\sun}$ -- $25 M_{\sun}$) with solar
composition. This study included the effects of rotation and angular
momentum transport with mass loss.  They found that the final specific
angular momentum distribution prior to core collapse had no dependence on
the initial surface velocity and only a weak dependence on the initial
mass; larger initial masses attain smaller specific angular momenta prior
to collapse because higher mass loss rates carry away more angular
momentum.  \citet{HLW00} also found that the $\Omega(r)$ profile in the
precollapse core is not a smooth function of radius, but consists of step
functions at fossil convective shell boundaries. The $15 M_{\sun}$ model
of \citet{HLW00} attains an angular velocity of 10 rad s$^{-1}$ (see their
Fig. 8) in the center of the iron core at the precollapse stage.  These
simulations did not include the effects of a magnetic field.  It is
possible that the iron core rotates more slowly if the effect of magnetic
braking is included \citep{spruit98,HW02}.  \citet{FH00} studied the
rotational effects on pure hydrodynamic core collapse explosions with
initial velocity profiles obtained by \citet{HLW00} with a central
rotational velocity of 4 rad s$^{-1}$.  We adopt the initial angular
velocity profile, $\Omega_{0}(r)$, of \citet{FH00} (hereafter called the
FH profile) as one case to study.

An analytical form of the $\Omega_{0}(r)$ profile had been adopted in some
previous work \citep{MM89, yamada94, FH00}.  One representation of the
profile takes the form \citep{MM89}: 
\begin{equation} 
\Omega_{0}(r) = \Omega_{\rm{0\_c}}\frac{R^2}{r^2 + R^2}, 
\end{equation}
where $R=10^{8}$ cm (hereafter called the MM profile).  The FH and MM
profiles with $\Omega_{\rm{0\_c}} = 4.0$ rad s$^{-1}$ resemble
each other \citep{FH00}, except that the MM profile is a continuous
function, and the FH profile is broken by steps.  The pioneering work on
core collapse supernova explosions of rotating stars by \citet{LW70}
employed an initial solid body rotation with 0.7 rad s$^{-1}$.  

\begin{figure}[htp]
\centering
\includegraphics[totalheight=0.8\textheight]{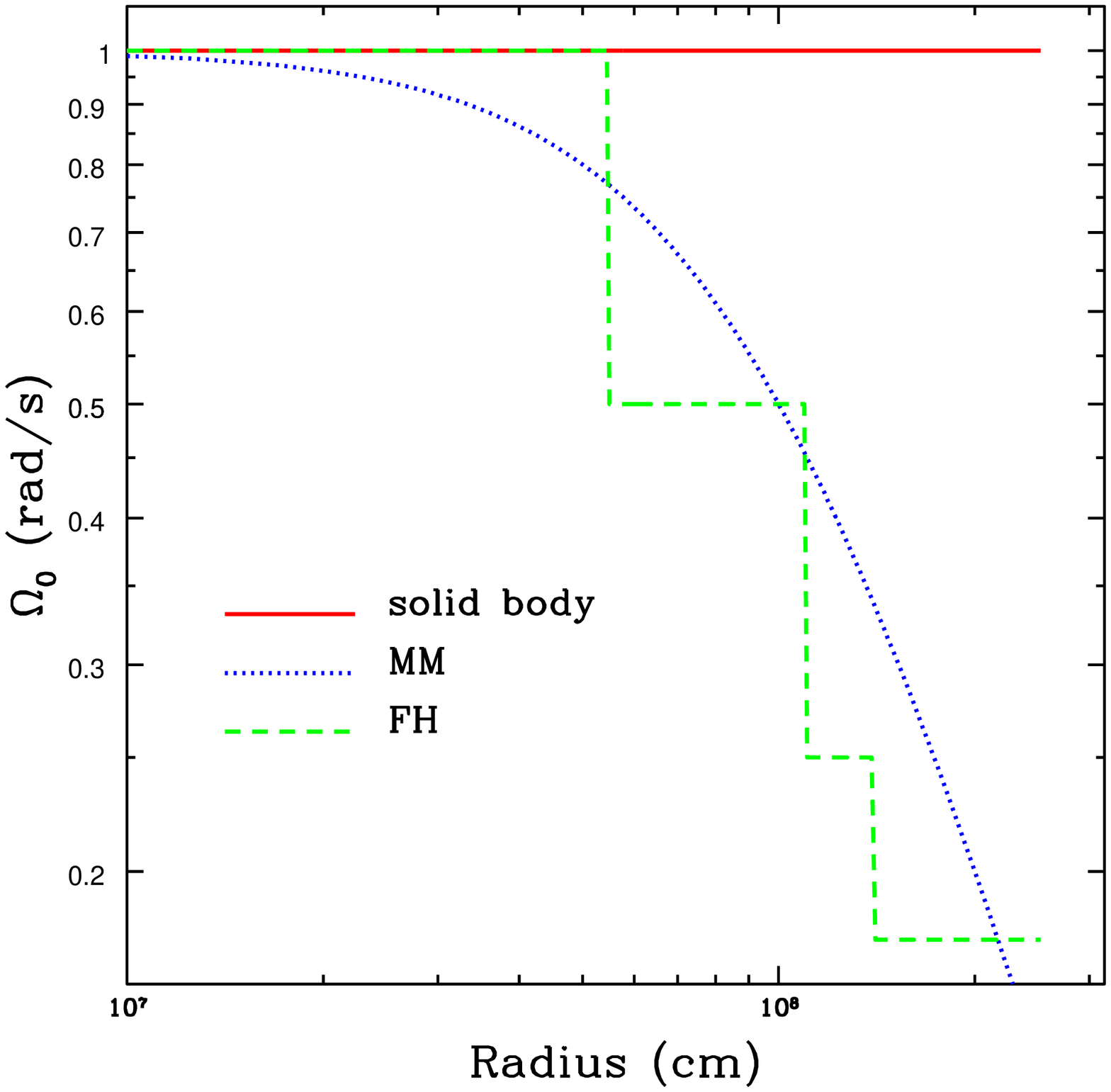}
\figcaption[omega0.eps]{Initial angular velocity profiles employed for
this study: a solid body rotational profile and two differential
rotational profiles, MM \citep{MM89} and FH \citep{FH00}.\label{rot0}}
\end{figure}

We adopt
the following three profiles for this study; solid body, MM, and FH
profiles, as illustrated in Fig. \ref{rot0}.  These profiles are scaled by
the central value of $\Omega_{\rm{0\_c}}$ that we choose to range from 0.2
to 10 rad s$^{-1}$.  As the collapse progresses, the PNS will rotate near
breakup speed for sufficiently high $\Omega_{0}(r)$.  The solid body case
is most susceptible to this problem.  For this work, we monitor the
rotation and omit cases that are not approximately self-consistent with our
spherically-symmetric treatment (although not necessarily unphysical). The
adopted profiles that we study in detail have small enough angular
momentum that little departure from spherical geometry will occur.

Once an $\Omega_{0}(r)$ profile is adopted, the $\Omega(r)$ profile is
computed using the radial density profiles produced by the core collapse
code described above.  We assume that the specific angular momentum is
conserved between consecutive time steps as follows:
\begin{eqnarray}
\label{omega1}
\Omega'(r') &=& \left( \frac{r}{r'} \right)^{2}\Omega(r) \\
\label{omega2}
&=& \left(\frac{\rho'}{\rho}\right)^{\frac{2}{3}}\Omega(r),
\end{eqnarray}
where the primed quantities represent the future time step
(the appropriate expressions for finite shells in difference
form was used in the code as applied below). 
It is inevitable that the collapsing core spins up and generates strong
differential rotation for very general choices of the $\Omega_{0}(r)$
profile, since the inner regions collapse larger relative distances than
the outer regions.  A strong shear must form at the boundary of the PNS.
At bounce, the original homologous core has a positive
gradient in $\Omega(r)$, but by 210 ms after bounce, this gradient has
nearly disappeared and after that the gradient is monotonically negative
outward (see below; Figs. \ref{difrot}, \ref{srot}).  

\begin{figure}[htp]
\centering
\includegraphics[totalheight=0.8\textheight]{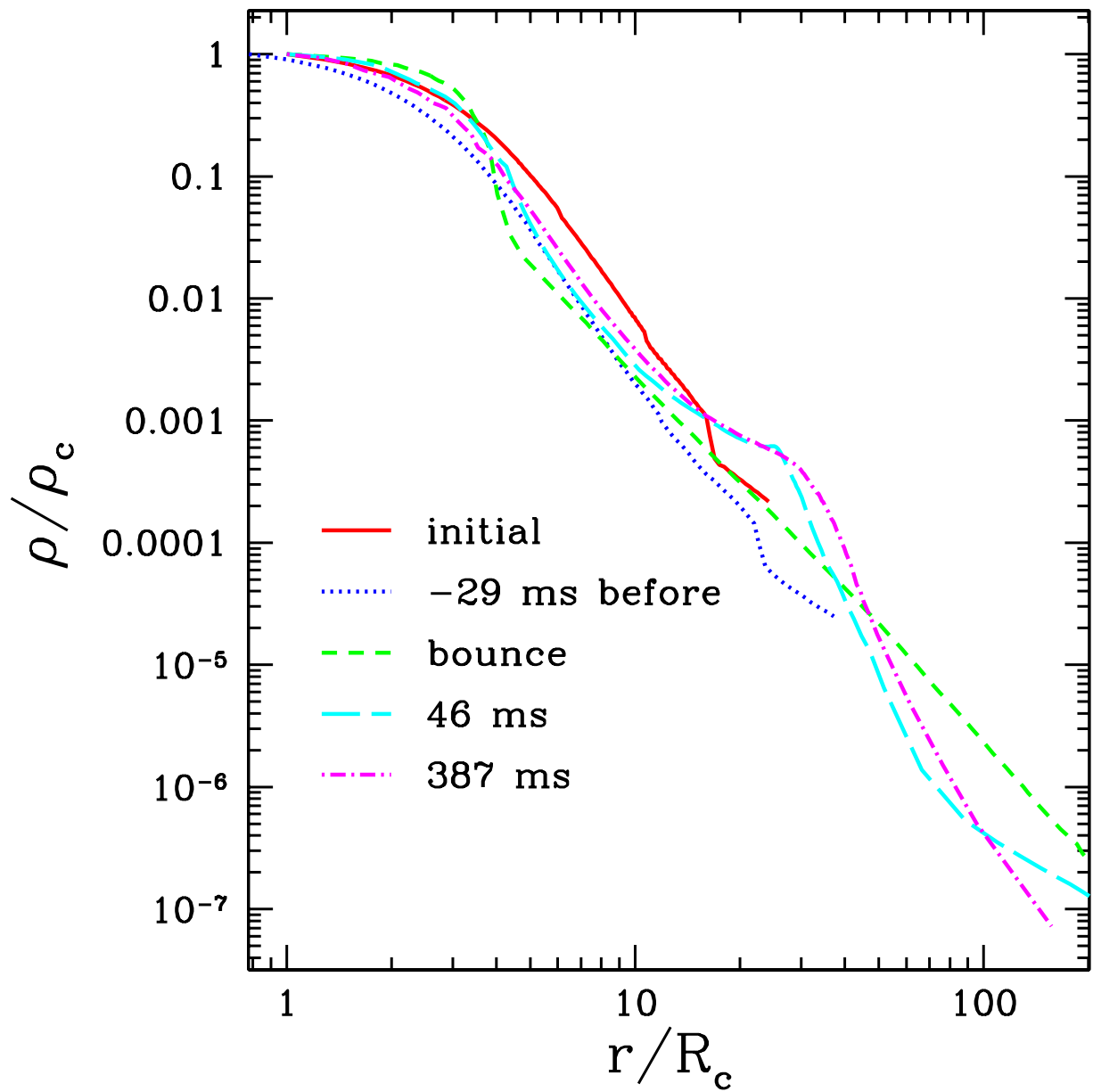}
\figcaption[densityr.eps]{The relative density profiles normalized to the 
central density for the initial iron core and for various
epochs in the collapse.  The central distribution at bounce is flatter
than the initial density profile resulting in a positive $\Omega$
gradient in the central regions, but the density profile relaxes to
being everywhere steeper than the initial profile after bounce leading
to a monotonically decreasing $\Omega$ as shown in Figs. \ref{difrot}
and \ref{srot}.\label{rho_norm}}   
\end{figure}

\citet{rud00} noted that the collapse of a white dwarf to a PNS gives 
a positive $\Omega(r)$ gradient since the relativistic degenerate 
core of the white dwarf has a steeper density profile than the PNS.
The PNS will thus be relatively more compact for a given central density.  
There are two important differences in the current calculations.  
The most critical is that we are not considering a core 
collapsing in isolation as for the accretion-induced collapse scenario.  
Rather, the PNS forms within the massive star collapse ambience, 
and the PNS must be strongly differentially
rotating with respect to the still-infalling matter.  This will generate
a strong shear at the boundary of the PNS that would not pertain to
a collapsing isolated white dwarf.  Another, more subtle, difference
is the equation of state.  Here we are considering the collapse of
a partially degenerate iron core to a PNS.  The equation of state
is not as different as for the highly relativistic white dwarf collapsing
to non-relativistic neutron star.  An examination of the relative density
profiles shown in Fig. \ref{rho_norm} reveals that the homologous core at 
bounce is somewhat less centrally condensed than the original iron core; 
this results in the temporary positive gradient in $\Omega$ within the
homologous core as shown below in Figs. \ref{difrot}, \ref{srot}.  About
50 ms after bounce, the density profile is nearly identical to
that of the initial iron core, giving the nearly flat rotation
profile shown below.  After that, the density profile becomes
somewhat more centrally condensed than the original iron core
and the rotation profile decreases monotonically outward even
deep within the PNS.

\subsection{Magnetic Field}

The MRI generates turbulence in a magnetized rotating fluid body that
amplifies the magnetic field and transfers angular momentum.  In this
paper we attempt, for the first time, to apply the physics of the MRI to
the core collapse supernova environment where the MRI criterion is broadly
satisfied.  The MRI should pertain in this environment and amplify the
magnetic field exponentially and perhaps, in turn, power MHD bi-polar
flow or jets.  Key
questions are the amplitude of the magnetic field and the effect on the
dynamics.  Those two questions are deeply related, but with the
limitations of this proof-of-principle study in mind, we will estimate the
strength of the magnetic field due to the MRI during the core collapse
process and postpone a study of feedback effects to a later work.

Ignoring entropy gradients for the moment, the condition for the
instability of the slow magnetosonic waves in a magnetized, differentially
rotating plasma is
\citep{BH91,BH98}: 
\begin{equation}
\frac{d\Omega^2}{d\ln{r}} + ({\bf{k \cdot v_{\rm{A}}}})^2 < 0,
\end{equation}
where 
\begin{equation}
\label{v_alfven}
v_{\rm{A}} = \frac{B}{\sqrt{4 \pi \rho}}
\end{equation}  
is the Alfv\'{e}n velocity.  When the magnetic field is very small, and/or 
the wavelength is very long,
$({\bf{k \cdot v_{\rm{A}}}})^2$ is negligible, and the instability
criterion for the MRI is simply that the angular velocity gradient be
negative \citep{BH91,BH98}, i.e.:
\begin{equation}
\frac{d\Omega^2}{d\ln{r}} < 0.
\end{equation}

The growth of the magnetic field associated with the MRI is exponential
with characteristic time scale of order $\Omega^{-1}$ since $\tau \sim 2
\pi/|\omega| \sim 2 \pi |d\Omega^2/d\ln{r}|^{-1/2}$. The time scale for 
the maximum growing mode is given by \citep{BH98}:
\begin{equation}
\label{growth-time}
\tau_{\rm{max}} = 4 \pi \left| \frac{d\Omega}{d\ln{r}} \right|^{-1}.
\end{equation}
We thus expect the MRI to dominate any process such as ``wrapping of field
lines'' (see Wheeler et al. 2000 and references therein) that only grows
linearly in time, even if on about the same time scale.  The MRI will also
operate under conditions of moderate rotation that are not sufficient to
compete with the PNS convective time scales to drive the sort of
$\alpha$~--~$\Omega$ dynamo invoked by, e.g., Duncan \& Thompson (1992).  
The resulting unstable flow is expected to become non-linear, develop
turbulence, and drive a dynamo that amplifies and sustains the field.  
The field will grow because of the MRI dynamo action until it reaches a
saturation field limit.  The details of the dynamo action due to MHD
turbulence are beyond the scope of this paper, and we ignore the effect of
the turbulence on the hydrodynamics.

\subsubsection{Saturation Fields}

An order of magnitude estimate for the saturation field can be obtained by
equating the shearing length scale $l_{\rm{shear}} \sim
dr/d\ln{\Omega}$ to the characteristic mode scale $\l_{\rm{mode}} \sim
v_{\rm{A}} \cdot (d\Omega/d\ln{r})^{-1}$.  The resulting saturation 
magnetic field is 
given by:
\begin{equation}
\label{bsat}
B_{\rm{sat}}^2 \sim 4 \pi \rho r^2 \Omega^2.
\end{equation}
This is the same result as obtained by setting the Alfv\'{e}n velocity
equal to the local rotational velocity, $v_{\rm{A}} = r \Omega$.  
In the subsequent discussion, we will adopt the field in eq. (\ref{bsat}) 
as the fiducial saturation field.  
Note that this expression does not depend on the 
shear, although the field growth in this case will depend on the shear
through the instability criterion.  Another simple estimate of the saturation 
field is to set the mode length equal to the local radius, on the grounds 
that the mode cannot be larger in wavelength than the physical size of the 
region in which it grows.  The resulting saturation field is: 
\begin{equation}
\label{br} B_{\rm{rad}}^2 \sim 4 \pi \rho r^2 \Omega^2 \left(
\frac{d\ln{\Omega}}{d\ln{r}}\right)^2.
\end{equation}
This differs from the previous estimate only by the logarithmic gradient
of $\Omega$, a factor normally of order unity. 

The empirical value of the saturation field obtained by the numerical 
simulation of \citet{HGB96} is: 
\begin{eqnarray} 
B_{\rm{sim}} &=& \sqrt{\frac{\rho}{\pi}} r \Omega \nonumber \\ 
\label{bsim}
&=& \frac{1}{2\pi} \cdot B_{\rm{sat}}.
\end{eqnarray}
This saturation field is achieved after turbulence is fully established,
which takes about 20 rotations following the initial exponential growth
\citep{HGB96}.  The field of eq. (\ref{bsim}) is another possible
saturation limit.  Note that for conditions of rotation at much less than
Keplerian, to which we restrict the current analysis, these saturation
fields and that to be described below, represent fields that are much 
less than the equipartition field, for which $B^2/8\pi$ is comparable to 
the ambient pressure, i.e. for the current calculations, 
$c_{\rm{s}} \gg r\Omega \sim v_{\rm{A}}$ (Fig. \ref{vel}).

Another means of estimating the saturation field considers the MRI mode
that grows at the maximum rate (eq. (\ref{growth-time})). When a vertical
field exists, the maximum unstable growing mode \citep{BH98} implies a
saturation field of:
\begin{equation}
\label{bmax}
B_{\rm{max}}^2 = - 4 \pi \rho \lambda_{\rm{max}}^2 \Omega^2 \cdot
\left[ \frac{1}{8 \pi^2} \left( 1 + \frac{1}{8} 
\frac{d\ln{\Omega^2}}{d\ln{r}} \right) 
\frac{d\ln{\Omega^2}}{d\ln{r}}\right], 
\end{equation}
where $\lambda_{\rm{max}}$ is the wavelength of the maximum growing mode.  
Once again, we argue that this fastest growing mode must be contained in
the physical region in which growth occurs, so the wavelength is not
allowed to exceed the local radius $r$.  With $\lambda_{\rm{max}} = r$,
eq. (\ref{bmax})  becomes:
\begin{equation}
\label{bmax2}
B_{\rm{max}}^2 = - B_{\rm{sat}}^2 \cdot \left[ \frac{1}{8 \pi^2} \left( 1
+ \frac{1}{8} \frac{d\ln{\Omega^2}}{d\ln{r}} \right)  
\frac{d\ln{\Omega^2}}{d\ln{r}}\right].
\end{equation}
This expression for the saturation field depends on the shear explicitly 
as well as indirectly through the stability criterion.
 Yet another variation would be to set the wavelength of the maximum
growing mode equal to the shear length, but all these variations just
yield differences of order $d\ln{\Omega}/d\ln{r}$.  Note that for the
maximum growing mode the expression for $B_{\rm{max}}^2$ acquires a
negative value when
\begin{eqnarray}
\label{omega-limits}
\frac{d\Omega^2}{d\ln{r}} & < & - 8\Omega^2 \mbox{ or }, \nonumber \\
\kappa^2 & < & - 4\Omega^2 < 0,
\end{eqnarray}
where $\kappa$ is the epicyclic frequency:
\begin{equation}
\kappa^2 = \frac{1}{r^3}\frac{d(r^4\Omega^2)}{dr} = 4\Omega^2 + 
\frac{d\Omega^2}{d\ln{r}}.
\end{equation} 
When eq. (\ref{omega-limits}) is true, the epicyclic motion dominates over
the MRI and prevents growth of the perturbation.  In practice, the
gradient of $\Omega$ may be reduced by mixing due to the epicyclic motion,
and the MRI may eventually be active in a region in which it was at first
suppressed by a strong negative gradient of $\Omega(r)$.  In this work,
we simply turned off field amplification when this condition arose,
thus, perhaps, minimizing the actual effect of the MRI.

In many cases, including our collapsing core, the background gas is unstable to
convection due to an entropy gradient.  The previous discussion can be
modified to take this effect into account.  When differentially rotating
gas is unstable to the MRI and stable to convection, a perturbed fluid
element experiences a stabilizing force due to the convective stability.  
Therefore, convective stability tends to stabilize the MRI.  The modified
instability criteria for the MRI with an entropy gradient are
\citep{BH91,BH98}:
\begin{eqnarray}
\label{crit1}
N^2 + \frac{\partial{\Omega^2}}{\partial{\ln{\varpi}}} < 0, \\
\label{crit2}
\left(-\frac{\partial{P}}{\partial{z}}\right)  
\left(\frac{\partial{\Omega^2}}{\partial{\omega}}\frac{\partial{s}}{\partial{z}} 
- \frac{\partial{\Omega^2}}{\partial{z}}\frac{\partial{s}}{\partial{\omega}}
\right) < 0,
\end{eqnarray}  
where $\varpi$ and $z$ are in cylindrical coordinates, s is entropy, and N 
is the Brunt-V\"{a}is\"{a}l\"{a} frequency:
\begin{equation}
\label{brunt}
N^2 = \frac{g\delta}{C_p}{\bigtriangledown}s
\end{equation}
where $\delta = -\left(\frac{\partial{\ln\rho}}{\partial{lnT}}
\right)_{P}$, $C_{P}$ is the specific heat at constant pressure.

For our spherical model, the LHS of eq. (\ref{crit2}) is zero.  
Note that the pure hydrodynamic counterparts to this instability, 
the H{\o}iland criteria, can be obtained
by simply replacing angular velocity with angular momentum, in
eq. (\ref{crit1}) and eq. (\ref{crit2}), but the H{\o}iland criteria
cannot be obtained by taking the limit of zero magnetic field
since eq. (\ref{crit1}) and eq. (\ref{crit2}) do not depend on field
strength \citep{BH98}.  
The spherically symmetric collapse models used in this work develop
regions that are unstable to convection, corresponding to a 
negative value of $N^2$, but are not able to compute the development of
convection.  The breakout of convection, although complex and time-dependent,
would tend to lower the ``superadiabatic" gradient and hence to
reduce the value of $|N^2|$ compared to the values we compute here from
eq. (\ref{brunt}).  For this reason, the values of $N^2$ we compute and
employ in eq. (\ref{crit1}) and other expressions may be exaggerated
in convectively unstable regions, with implications for our calculations 
that will be described below. 

For a rotating magnetized object with axial symmetry, the maximum 
growing mode in the presence of entropy gradients
corresponding to $\lambda_{\rm{max}} = r$ is:
\begin{equation}
\label{bmaxentropy}
B_{\rm{max,en}}^2 = - B_{\rm{sat}}^2 \cdot \left[ \frac{\eta^2 -
1}{4\pi^2} + \frac{1}{8\pi^2} \left( \eta + \frac{1}{8} \left( \xi
\frac{N^2}{\Omega^2} + \eta \frac{d\ln{\Omega^2}}{d\ln{r}} \right)  
\right) \left( \xi \frac{N^2}{\Omega^2} + \eta
\frac{d\ln{\Omega^2}}{d\ln{r}} \right) \right],
\end{equation}
and the growth time scale is given by:
\begin{equation}
\tau_{\rm{max,en}} = 2 \pi \left|(\eta^2 - 2\eta + 1)\Omega^2 +
\frac{\eta -1}{2}\left( \xi N^2 + \eta \frac{d\Omega^2}{d\ln{r}}
\right) + \frac{1}{16\Omega^2} \left(\xi N^2 + \eta
\frac{d\Omega^2}{d\ln{r}}\right)^2 \right|^{-\frac{1}{2}},
\end{equation}
where:
\begin{eqnarray}
\xi^2 & = & (1-\sin 2\theta)^2, \\
\eta^2 & = & \sin^2 \theta (1-\sin 2\theta).
\end{eqnarray}

When the polar angle $\theta = \pi/2$, we obtain the instability criterion 
in the equatorial plane:
\begin{equation}
\label{Nstability}
N^2 + \frac{d\Omega^2}{d\ln{r}} < 0,
\end{equation}
and the saturation field of the maximum growing mode becomes:
\begin{equation}
\label{bmaxentropy2}
B_{\rm{max,en}}^2 = - B_{\rm{sat}}^2 \cdot \left[ \frac{1}{8 \pi^2} \left(
1 + \frac{1}{8} \left( \frac{N^2}{\Omega^2} +
\frac{d\ln{\Omega^2}}{d\ln{r}} \right) \right) \left( \frac{N^2}{\Omega^2}
+ \frac{d\ln{\Omega^2}}{d\ln{r}} \right) \right],
\end{equation} 
which grows exponentially with the time scale
\begin{equation}
\label{growth-time2}
\tau_{\rm{max,en}} = 4\pi \left| \frac{N^2}{2\Omega} + 
\frac{d\Omega}{d\ln{r}} \right|^{-1}.
\end{equation}
The field growth for the case of the saturation field 
in eq. (\ref{bmaxentropy2}) depends explicitly on the shear and 
Brunt-V\"{a}is\"{a}l\"{a} frequency and depends indirectly on those quantities
through the instability criterion and growth time.
For $N^2 = 0$, the saturation field of the maximum growing mode and its
growth time scale reduce to those of the no--entropy--gradient case (eq.
(\ref{growth-time}), (\ref{bmax2})).  

From eq. (\ref{bmaxentropy2}),
$B_{\rm{max,en}}^2$ becomes negative due to competition with the
epicyclic frequency when
\begin{equation}
\label{negativeb}
N^2 + \frac{d\Omega^2}{d\ln{r}} < -8\Omega^2.
\end{equation}
Convective instability ($N^2 < 0$) assists the MRI to destabilize the
flow according to the instability criterion in eq. (\ref{Nstability}).  
For the maximum growing mode, however, even when the instability criterion 
is met, the growth of the mode is inhibited when the magnitude of 
$N^2$ is negative and large because of the condition in 
eq. (\ref{negativeb}).  In our calculations, the field with saturation 
limit $B_{\rm{sat}}$ grows whenever the instability criterion is 
satisfied.  In this case, the MRI is inhibited in regions of convective
stability and promoted in regions of convective instability.  The field 
with saturation limit $B_{\rm{max,en}}$ grows when the instability
criterion is satisfied, but epicyclic motions do not dominate, i. e.,
when $-8\Omega^2 < N^2 + d\Omega^2/d\ln{r} < 0$.
Note that our estimates of $N^2$ in convectively unstable regions
may exaggerate the effects of convection in eqs. (\ref{Nstability}),
(\ref{bmaxentropy2}), (\ref{growth-time2}), and (\ref{negativeb}).
In particular, an exessively large negative value of $N^2$ in
eq. (\ref{Nstability}) could promote the instability to the MRI
artificially.  This could affect either prescription for saturation
fields.  On the other hand, an excessively large negative value for
$N^2$ in eq. (\ref{negativeb}) will inhibit the MRI from operating in 
conditions where it might actually do so by exaggerating the epicyclic 
effects.  In this sense, our estimates of the activity of the MRI may 
be conservative.

In the numerical work to be presented below, we considered two
prescriptions for the saturation field in the equatorial plane, the
fiducial field $B_{\rm{sat}}$ corresponding to $v_{\rm{A}} = r\Omega$ (eq.  
(\ref{bsat})), and $B_{\rm{max,en}}$, which corresponds to the maximum
growing mode with the effects of the entropy gradient included (eq.
(\ref{bmaxentropy2})).  We apply the instability criterion, 
eq. (\ref{Nstability}), and the growth time of eq. (\ref{growth-time2})
to both saturation field prescriptions and the epicyclic limit,
eq. (\ref{negativeb}), to the prescription of eq. (\ref{bmaxentropy2}).
Cases corresponding to the numerical saturation limit (eq. (\ref{bsim})) 
can be obtained by simply dividing the resulting field profile in the 
fiducial saturation field case by 2$\pi$.

Any initial vertical field is turned into a radial and an azimuthal
field by the MRI, leading to channel flows in 2D simulations and spiral
streaming in 3D simulations \citep{BH98}.  Other initial field
configurations are also known to be unstable \citep{VD92,BH98}.  The flow
induced by the MRI quickly develops turbulence. \citet{HGB96} found in
their study of dynamo action due to the MRI that the saturation magnetic
energy has a very weak dependency on the initial field strength and
configuration as long as there is some vertical component to the original
seed field.  Both the field strength and the configuration of the magnetic
field in the precollapse iron core are unknown.  Here, we assume that in
the precollapse iron core the magnetic field is much weaker than the
equipartition field so that the MRI is strongly unstable, resulting in
exponential growth of the field, and that there is some vertical component
to the field so that the saturation fields we adopt are representative of
the strongest limit expected from the MRI.  These preliminary calculations
represent a reasonable approximation of the effect of the MRI on growth of
the magnetic field in the rotating and collapsing cores that precede the
supernovae explosions.

\subsubsection{Time Dependent Calculation}

We applied the time scale, $\tau$, of eq. (\ref{growth-time2}) 
so that the magnetic
field amplified by the MRI achieves the asymptotic saturation value
$B_{\rm{asym}}$ only on this time scale.  To account for this finite
growth time to reach saturation, we take the growth of the magnetic field
$\Delta B$ during interval $\Delta t$ to be:
\begin{equation}
\label{delb}
\Delta B = (1 - e^{-\frac{\Delta t}{\tau}})(B_{\rm{asym}} - B_{\rm{old}}),
\end{equation}
where B$_{\rm{asym}}$ is calculated using eq. (\ref{bsat}) or eq.  
(\ref{bmaxentropy2}) and $B_{\rm{old}}$ is the magnetic field achieved in
the previous time step.  Note that the growth time depends on $N^2$
and may be artificially short in strongly convectively unstable regions.
The magnetic field for the current time step is then computed as:
\begin{equation}
\label{delb2}
B = B_{\rm{old}} + \Delta B,
\end{equation}
where $\Delta$B goes to zero as the field  asymptotically approaches the 
saturation limit.

\subsection{MHD Power}

MHD jets are common in systems with a central object that is accreting
matter with angular momentum and a magnetic field \citep{meier01}.  
Examples of such systems are the engines of active galactic nuclei, micro
quasars, gamma--ray bursts (collapsars), and young stellar objects.  In
the collapse of a rotating iron core, the environment for MHD jets to form
is satisfied by a central PNS with infalling matter with angular momentum
and a magnetic field.  Previous studies have shown that buoyant MHD
outflows form in the collapse of a magnetized rotating iron core
\citep{LW70,sym84}.  \citet{wheel02} discussed jet formation in the
context of magnetorotational core collapse supernovae, but based only on
the linear wrapping of field lines.  Although jet formation is beyond the
scope of this paper, we can estimate the power produced by the magnetic
field generated by the MRI that could drive bi-polar flows or jets.

The characteristic power of non-relativistic MHD outflow is given by
\citet{blan82}; see also \citep{meier99,wheel02}:
\begin{equation}
\label{lmhd}
L_{\rm{MHD}} = \frac{B^2r^3\Omega}{2}. 
\end{equation}
The outflow carries energy, angular momentum, and mass.  The ejected mass is
contained in a cone with a fractional solid angle $f_{\Delta \Omega} =
\Delta \Omega/4\pi$, where $\Delta \Omega$ is the solid angle of the cone
about the rotational axis of the PNS.  The critical power required for the
plasma in the cone to escape from the gravitational potential during a
free fall time is given by \citet{meier99} and \citet{wheel02},
\begin{equation}
L_{\rm{crit}} = \frac{E_{\rm{esc}}}{\tau_{\rm{ff}}} = 
\frac{4 \pi}{3} \rho r^2 f_{\Delta \Omega} 
\left(\frac{GM}{r}\right)^{\frac{3}{2}}.
\end{equation}
When $L_{\rm{MHD}}$ exceeds $L_{\rm{crit}}$, the plasma can be accelerated
to escape velocity or higher.  The ratio of $L_{\rm{MHD}}$ to
$L_{\rm{crit}}$ is expressed as a dimensionless parameter,
\begin{equation}
\label{nu}
\nu \equiv \left(\frac{L_{\rm{MHD}}}{L_{\rm{crit}}}\right)^{\frac{1}{2}} = 
f^{-1/2}_{\Delta \Omega}
\left(\frac{3\Omega}{\Omega_{\rm{kep}}}\right)^{\frac{1}{2}} 
\left(\frac{v_{\rm{A}}}{v_{\rm{esc}}}\right).
\end{equation}
Numerical calculations have shown that when $\nu$ is less than 1, 
the dynamics correspond to a Blandford-Payne-type outflow 
\citep{meier97,meier99}, but when $\nu$ is greater than 1, 
the character of the flow changes to that of a faster
and more tightly collimated jet.  For our case, we expect a
non-relativistic Blandford-Payne-type outflow, which has a broad opening
at the base of the jet.  For simplicity, we assume $f_{\Delta \Omega}=1$.

\section{Results}

\begin{figure}[htp]
\centering
\includegraphics[totalheight=0.8\textheight]{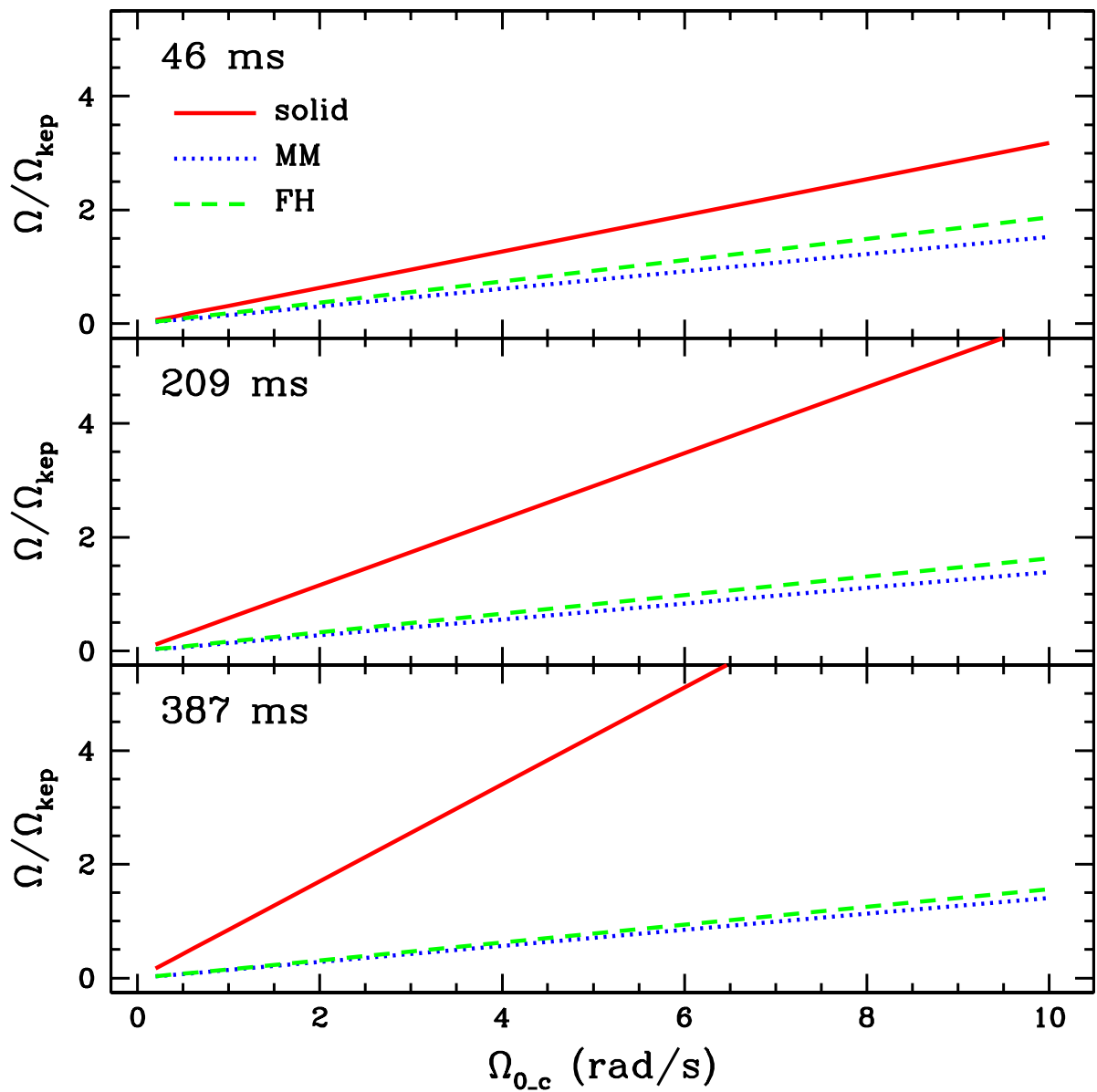}
\figcaption[omegakepwin.eps]{Evolution of $\Omega/\Omega_{\rm{kep}}$ as a
function of initial central angular velocity for each initial rotational
profile.  We restrict our analysis to $\Omega_{\rm{0\_c}} < 0.2$ rad
s$^{-1}$ for the initial solid body, and $\Omega_{\rm{0\_c}} < 1.0$ rad
s$^{-1}$ for the initial differential rotational profiles (MM and FH) in
order to justify our spherical approximation.\label{omegakep}}
\end{figure}

Because the code used for these calculations employs spherical geometry
and cannot include rotational effects explicitly, our study is limited to
the ``slow rotation'' case, where we can assume spherical geometry to good
approximation.  The initial rotational velocities in the iron core are so
small that a significant departure from spherical geometry is not expected
except for the solid rotation cases with $\Omega_{\rm{0\_c}} > 2.0$ rad
s$^{-1}$.  As the core collapses and density increases, however, it is
inevitable that the rotational velocity increases (eq. (\ref{omega1}),
(\ref{omega2})) while generating strong differential rotation, even for
initially slow rotation.  Fig. \ref{omegakep} shows the time evolution of
the maximum value of the ratio of $\Omega(r)$ to the local Keplerian
velocity as a function of $\Omega_{\rm{0\_c}}$.  For all initial
rotational profiles, the ratio is linear with $\Omega_{\rm{0\_c}}$ at a
given time.  At 46 ms after bounce, some material is already rotating
above break-up speed if $\Omega_{\rm{0\_c}} \gta $ 5.5 rad s$^{-1}$ for
the MM and FH initial profiles and if $\Omega_{\rm{0\_c}} \gta$ 3.0 rad
s$^{-1}$ for the solid initial profile.  As time elapses, the solid
profile case acquires break-up velocity for lower $\Omega_{\rm{0\_c}}$;
$\Omega/\Omega_{\rm{kep}} \sim$ 1 for $\Omega_{\rm{0\_c}} \sim$ 1.2 at 387
ms after bounce. Therefore for $\Omega_{\rm{0\_c}} \gta 1.2$, the envelope
is expected to be severely distorted, possibly creating a disk-like
structure.  The initial differential rotation cases with the MM and FH
profiles behave differently from the solid profile.  The maximum value of
the ratio $\Omega/\Omega_{\rm{kep}}$ decreases with time.  To be
self-consistent, we will analyze only the initial solid profile with
$\Omega_{\rm{0\_c}} \leq 0.2$ rad s$^{-1}$, and the MM and FH profiles
with $\Omega_{\rm{0\_c}} \leq 1.0$ rad s$^{-1}$.  We stress that more
rapid rotation in the initial models is not necessarily unphysical, only
that we cannot treat such cases here.  By selecting slow rotation cases,
the structure is assumed to be stable to nonaxisymmetric instabilities.

\begin{figure}[htp]
\centering  
\includegraphics[totalheight=0.8\textheight]{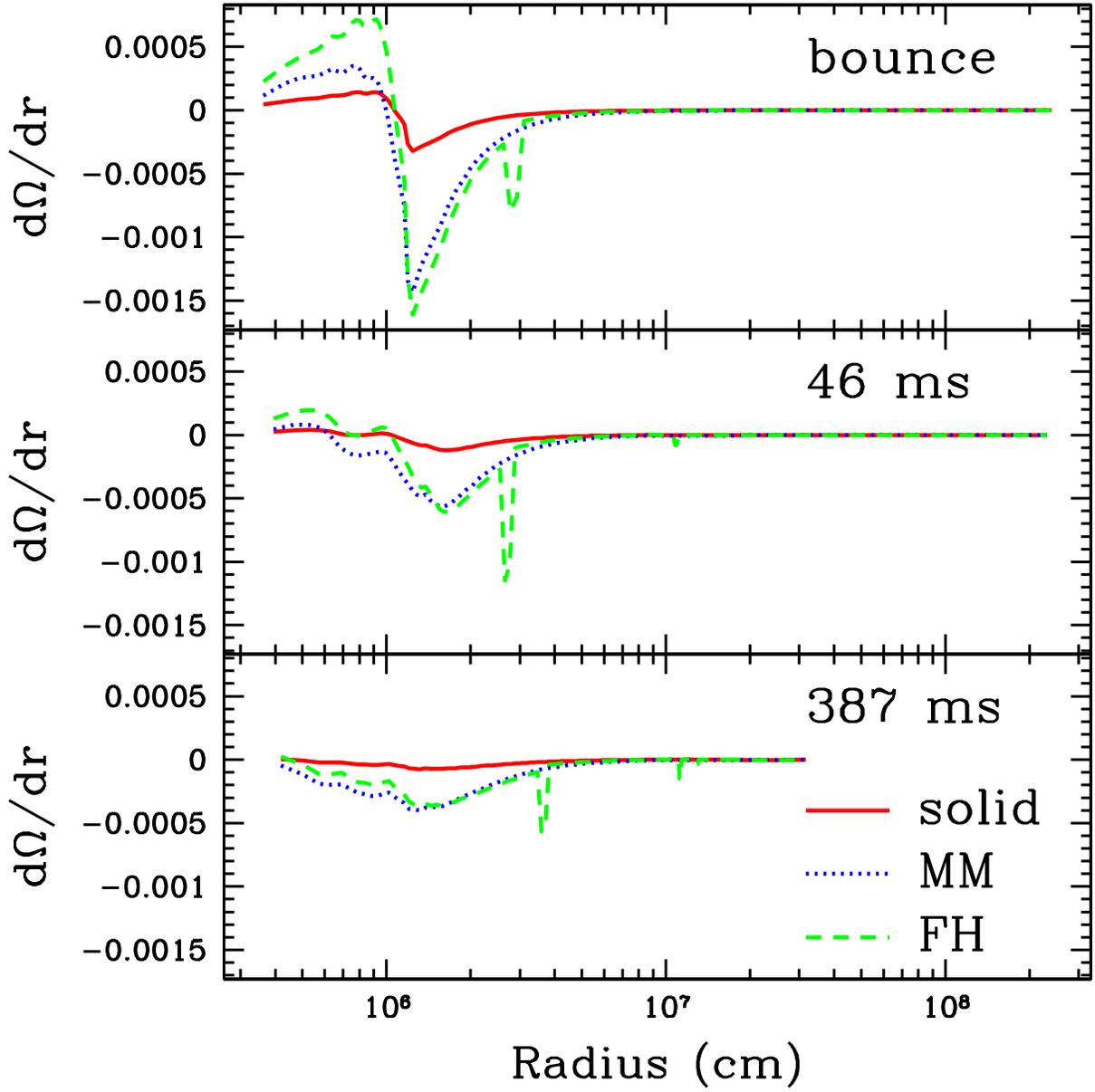}
\figcaption[shear.eps]{Plot of shear for a solid body profile and
differential rotational profiles (MM and FH) at bounce and at 46 ms and
387 ms after bounce.\label{shear}}
\end{figure}

As expected, the shear is the strongest at the boundary of the initial
homologous core as seen in Fig. \ref{shear}.  At bounce the shear is
positive inside of the initial homologous core, and the region is stable
against the MRI though convective instability can destabilize the
structure.  The solid body profile possesses a similar shear profile to
the FH and MM profiles, but the magnitude of the shear is smaller since
this case started with smaller $\Omega_{\rm{0\_c}}$.  The spikes in the FH
profile are features due to the discontinuities in $\Omega_0$.  By the end
of the calculations, the shear is negative in most regions, and the magnitude
of the shear is diminished for all cases.

\begin{figure}[htp]
\centering
\includegraphics[totalheight=0.8\textheight]{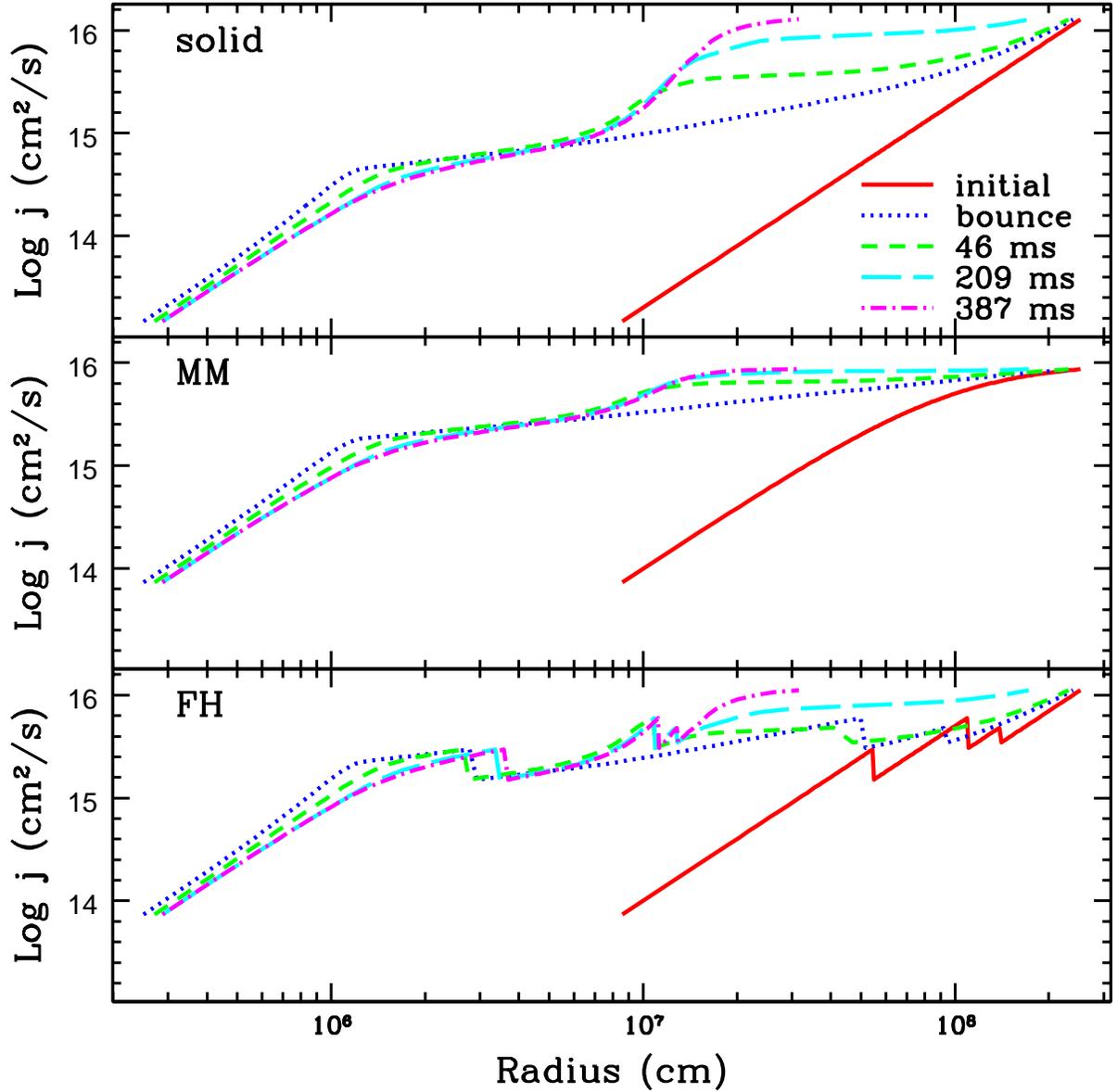}
\figcaption[momentum.eps]{Evolution of specific angular momentum
distribution for each initial rotational profile.  The gradient is
positive, except at discontinuities in the FH profile, indicating that the
structure is stable to the Rayleigh criterion.\label{j}}
\end{figure}

Fig. \ref{j} gives the evolution of the angular momentum per unit mass,
$j$, for representative initial models.  The structure is everywhere
stable to the Rayleigh criterion except at the composition discontinuities
in the FH model.  The solid body model with $\Omega_{\rm{0\_c}} = 0.2$ rad
s$^{-1}$ has smaller $j$ in the initial model and in the post-collapse
inner layers than the MM or FH cases with $\Omega_{\rm{0\_c}} = 1.0$ rad
s$^{-1}$ but has comparable values of $j$ in the outer regions at later
times.  

\begin{figure}[htp]
\centering
\includegraphics[totalheight=0.8\textheight]{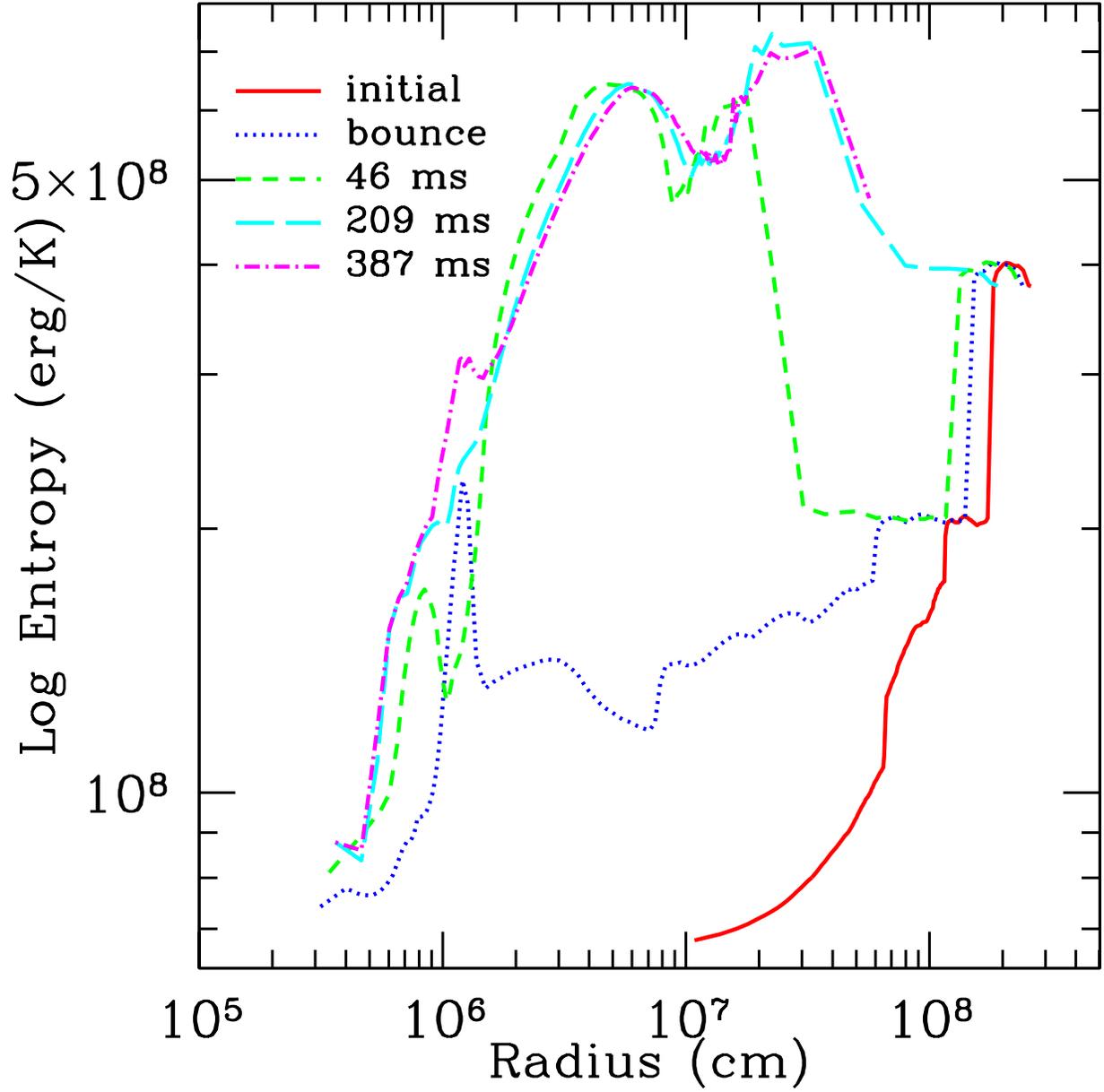}
\figcaption[entropy.eps]{Evolution of entropy in the collapsing core.  
Most regions between the boundary of the PNS and the stalled shock are
convectively unstable.\label{entropy}}
\end{figure}

The collapsing core has convectively stable regions with $N^{2} > 0$
and unstable regions with negative entropy gradient or $N^{2} < 0$ 
(Fig. \ref{entropy}).  In general, the PNS (less than 80 -- 90 km) is 
convectively stable, and the region between the boundary of the PNS and 
the stalled shock is convectively unstable. After bounce, 
the convective stability in the original homologous core
is sufficient to damp the MRI within the
PNS in accord with eq. (\ref{Nstability}).  For this
reason, the field within 10 km presented below was amplified before
bounce and then frozen in.  When ambient gas is unstable to convection, 
the MRI is enhanced (eq. (\ref{Nstability}), (\ref{bmaxentropy2}), 
(\ref{growth-time2})), unless the epicyclic frequency is too high 
(eq. (\ref{negativeb})) for the case with  $B_{\rm{max,en}}$.  
Note that this one-dimensional code does not 
actually include convection, so the Brunt-V\"{a}is\"{a}l\"{a} frequency we
compute here is a formal quantity and likely to be higher than the eddy 
overturn frequency of fully developed convection.  For this reason, the 
MRI may be artificially activated in convectively unstable regions with 
positive gradient of $\Omega$.  In practice, this does not appear to
be a significant factor since the structure is broadly unstable to
the MRI anyway.  The growth time in convectively unstable regions
could be artificially short.  The most significant effect of our
prescription for the Brunt-V\"{a}is\"{a}l\"{a} frequency may be to exaggerate 
the epicyclic effects through eq. (\ref{negativeb}), especially for
regions with small rotational velocity.  The result is to shut
off the MRI in regions with large negative $N^2$ and small $\Omega$.  This
has a substantial effect on the case with saturation field corresponding
to the maximum growing mode for which eq. (\ref{negativeb}) applies,
but does not affect the case with the fiducial saturation field.   


\subsection{Initial Differential Rotation Case}

\begin{figure}[htp]
\centering
\includegraphics[totalheight=0.8\textheight]{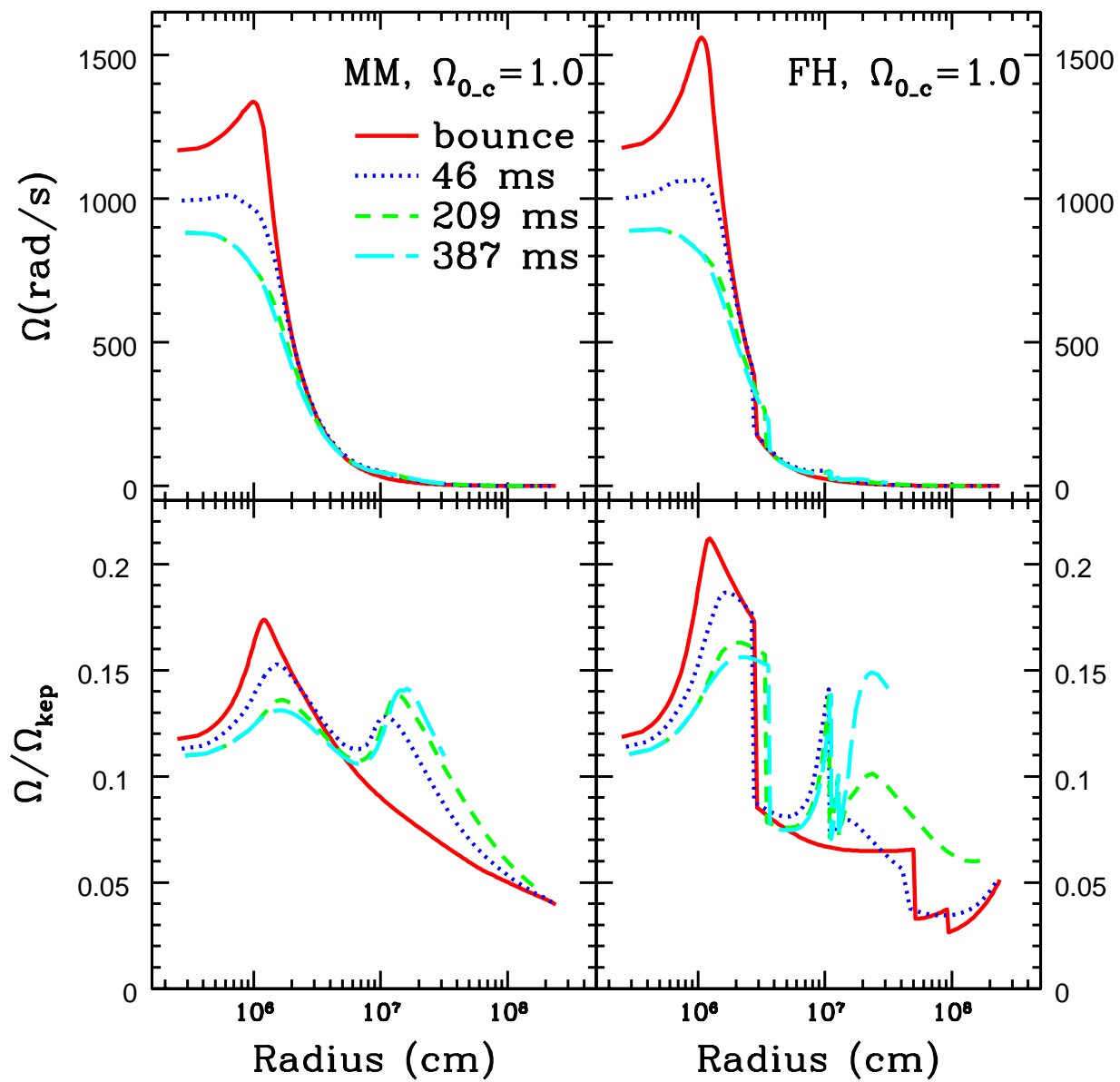}
\figcaption[omega1.eps]{Rotational profiles and $\Omega/\Omega_{\rm{kep}}$
for the initial differential rotation cases (MM and FH) with
$\Omega_{\rm{0\_c}} = 1.0$ rad s$^{-1}$.  The collapse generates strong
differential rotation.\label{difrot}}
\end{figure}

Both of our initial differential rotation profiles, FH and MM, produce
strong differential rotation (Fig. \ref{difrot}) due to collapse of the
core with the strongest shear at the boundary of the initial homologous
core (Fig. \ref{shear}) which later becomes the hydrostatic PNS core.  
At bounce, the initial homologous core has positive $\Omega(r)$ gradient, 
but the bounce of the initial homologous core eventually flattens the profile 
and makes the gradient negative.  The differences between the FH and MM 
profiles are that the former preserves discontinuities present in the initial 
profile in subsequent rotational profiles, as clearly seen in the plot of
$\Omega/\Omega_{\rm{kep}}$ in Fig. \ref{difrot}, and that the peak value
of $\Omega(r)$ at bounce in the FH profile is higher than that of the MM
profile even though they have the same $\Omega_{\rm{0\_c}} = 1.0$ rad
s$^{-1}$.  This higher maximum value of $\Omega$ at bounce is because 
the first step in the FH profile, which extends
to 5.3 $\times10^7$ cm in the initial iron core, gives a larger
$\Omega_{0}$ at finite radius than the MM profile (Fig. \ref{rot0}).  
Since both have the same
$\Omega_{\rm{0\_c}}$, the central rotational velocities stay the same:
1180 rad s$^{-1}$ at bounce and 890 rad s$^{-1}$ 387 ms after bounce,
which gives periods of 5.3 ms and 7.1 ms respectively.  Even a relatively
modest value of $\Omega_{0}$ gives a very rapidly rotating PNS
and hence strong differential rotation with respect to the infalling matter.

Because we have selected ``slow rotator'' conditions, the rotation remains
sub--Keplerian (Fig. \ref{difrot}).  At bounce, the peak of
$\Omega/\Omega_{\rm{kep}}$ is at the boundary of the initial homologous
core for both profiles, but the peak for the FH case has a higher value 
reflecting faster rotation for the reason mentioned above.  At later times, 
however, the peaks move to the second hump which is located inside the stalled
shock.  This hump is at the same location as a maximum in entropy (Fig.
\ref{entropy}) which is caused by shocked material with higher density.

\begin{figure}[htp]
\centering
\includegraphics[totalheight=0.8\textheight]{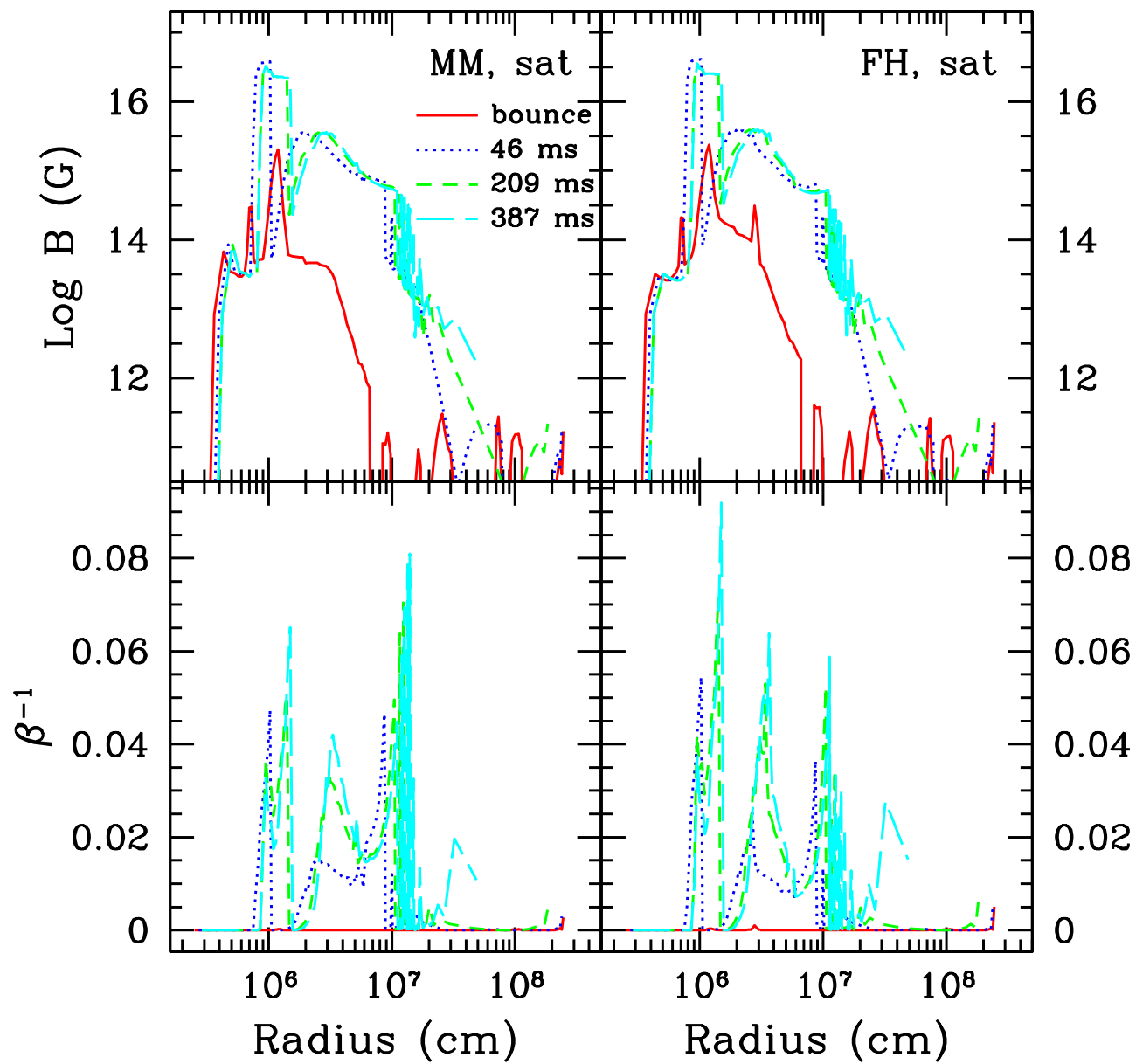}
\figcaption[brot.eps]{The magnetic field distribution and $\beta^{-1}$ for the
fiducial saturation field, $B_{\rm{sat}}$, for initial differential rotation
cases (MM and FH).  The maximum field is of order $\sim 10^{16}$ G by 387 ms
after bounce and is sub--equipartition.\label{brotdiff}}
\end{figure}

\begin{figure}[htp]
\centering
\includegraphics[totalheight=0.8\textheight]{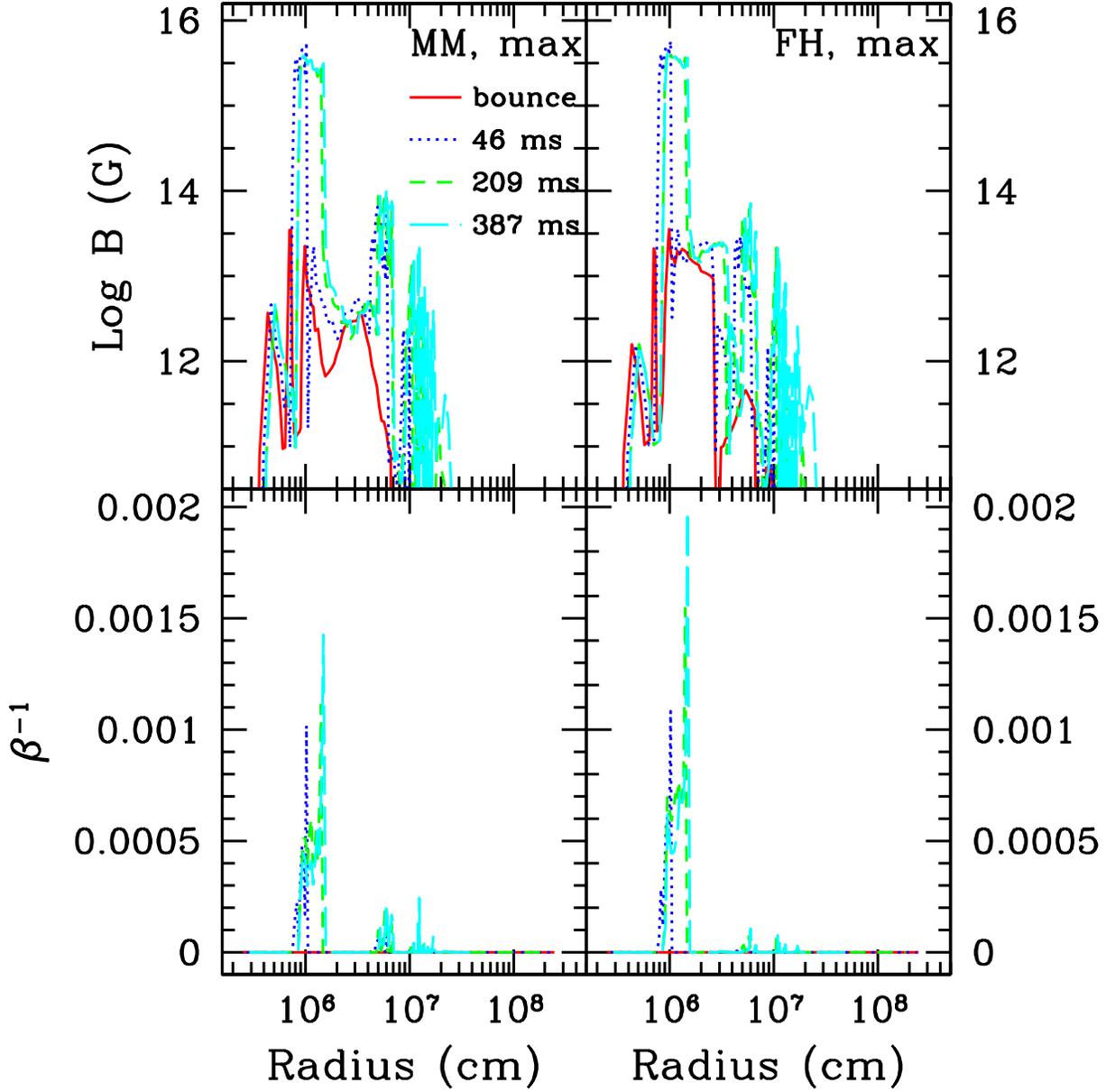}
\figcaption[bmax.eps]{The magnetic field distribution and $\beta^{-1}$ that
correspond to the saturation field of the maximum growing mode of the MRI,
$B_{\rm{max,en}}$, for the initial differential rotation cases (MM and FH).  
The maximum field is of order $\sim 10^{15}$ G by 387 ms after bounce and is
sub--equipartition.\label{bmaxdiff}}
\end{figure}

For the given initial rotational profiles, the fiducial field
$B_{\rm{sat}}$ and that corresponding to the maximum growing mode with
entropy gradient $B_{\rm{max,en}}$ (eq. (\ref{bmaxentropy2})) are
amplified exponentially according to eq. (\ref{delb}) and (\ref{delb2}).  
The resulting magnetic field and the ratio $\beta^{-1} \equiv
P_{\rm{mag}}/P_{\rm{gas}}$ (where $\beta$ is the conventional $\beta$ in plasma
physics) for $B_{\rm{sat}}$ are presented in Fig. \ref{brotdiff} and for
$B_{\rm{max,en}}$ in Fig. \ref{bmaxdiff}.  For both magnetic field
prescriptions, there is no significant difference between the MM and FH
profiles, except that FH achieves a slightly higher magnetic field and
thus larger $\beta^{-1}$ because of the higher rotational velocity at the
boundary of the initial homologous core where the shear is the strongest.  

Considerable differences occur between the $B_{\rm{sat}}$ and 
$B_{\rm{max,en}}$ cases.  These differences are due to the fact that 
the  $B_{\rm{sat}}$ case depends directly only on the azimuthal
angular velocity and not the shear.  This case  depends on the radial gradients
reflected in the shear and the Brunt-V\"{a}is\"{a}l\"{a} frequency only 
indirectly through the instability criterion, eq. (\ref{Nstability}),
and the growth time, eq. (\ref{growth-time2}).  The case with 
$B_{\rm{max,en}}$ depends both explicitly and implicitly on the 
angular velocity, the shear, and the Brunt-V\"{a}is\"{a}l\"{a} frequency 
through eqs. (\ref{Nstability}), (\ref{bmaxentropy2}), (\ref{growth-time2}),
and (\ref{negativeb}).  It is difficult to tell what the appropriate 
saturation field should be.  Given the limitation of the current 
calculations, we can only argue that these fields are roughly representative 
of what one would expect during core collapse.  We return to this issue 
in \S 4.

In the inner region (less than 7 km), for the $B_{\rm{sat}}$ case the
field was amplified before bounce and then remained constant after bounce
since the inner portions are sufficiently stable to convection to damp the
MRI there.  Near the boundary of the PNS, convective instability promotes
the MRI giving the strong flat peak at $\sim 10^{6}$ cm in Fig. 8.  The
sharp decline on the outer boundary of the peak is due to damping by
convective stability that diminishes in strength at larger radii, yielding
a gradual radial increase in field strength at later time.  For the
$B_{\rm{max,en}}$ case, the instability criterion is fulfilled before
bounce, but at the same time the condition in eq. (\ref{negativeb}) can be
met: the epicyclic motion prevents growth of the MRI, and in our
calculation the field growth is simply turned off while this condition
holds.  The result is smaller fields in the inner region for this case.  
This suppression of the fields for this case may be exaggerated by our
prescription for $N^2$ in the absence of actual convective motion.  In the
outer region near the shock front, the angular velocity gradient is mostly
negative and the shear is small. When ambient gas is convectively unstable
the MRI amplifies the magnetic field for the $B_{\rm{sat}}$ case.  For the
$B_{\rm{max,en}}$ case, however, the condition in eq. (\ref{negativeb}) is
met, again resulting in no amplification of the magnetic field.  This
interplay with convective instability results in the ``spikiness" of the
resulting field distribution.  Once again, this suppression of the MRI may
be artificial in these calculations. Some of the spikiness is exacerbated
by evaluating gradients simply from zone to zone rather than attempting
any smoothing.  We have done calculations in which we have formally set $N
= 0$.  The result is a substantially smooth distribution of fields at late
times that resembles the upper envelope of the field distribution given in
Fig. \ref{brotdiff}.

The $\beta^{-1}$ profiles have multiple peaks for the $B_{\rm{sat}}$ cases
but has single peak for the $B_{\rm{max,en}}$ case.  For the $B_{\rm{sat}}$
case, MHD activity is expected not only around the PNS boundary, but also
the region between the boundary and the stalled shock.  Note that at later
times $B_{\rm{sat}}$ is larger than $B_{\rm{max,en}}$ for both the FH and
MM profiles, and consequently $\beta^{-1}$ is larger for the $B_{\rm{sat}}$
case.  The parameter $\beta^{-1}$ is especially sensitive to the amplitude
of the magnetic field because it is proportional to the square of the
magnetic field.

\begin{figure}[htp]
\centering
\includegraphics[totalheight=0.8\textheight]{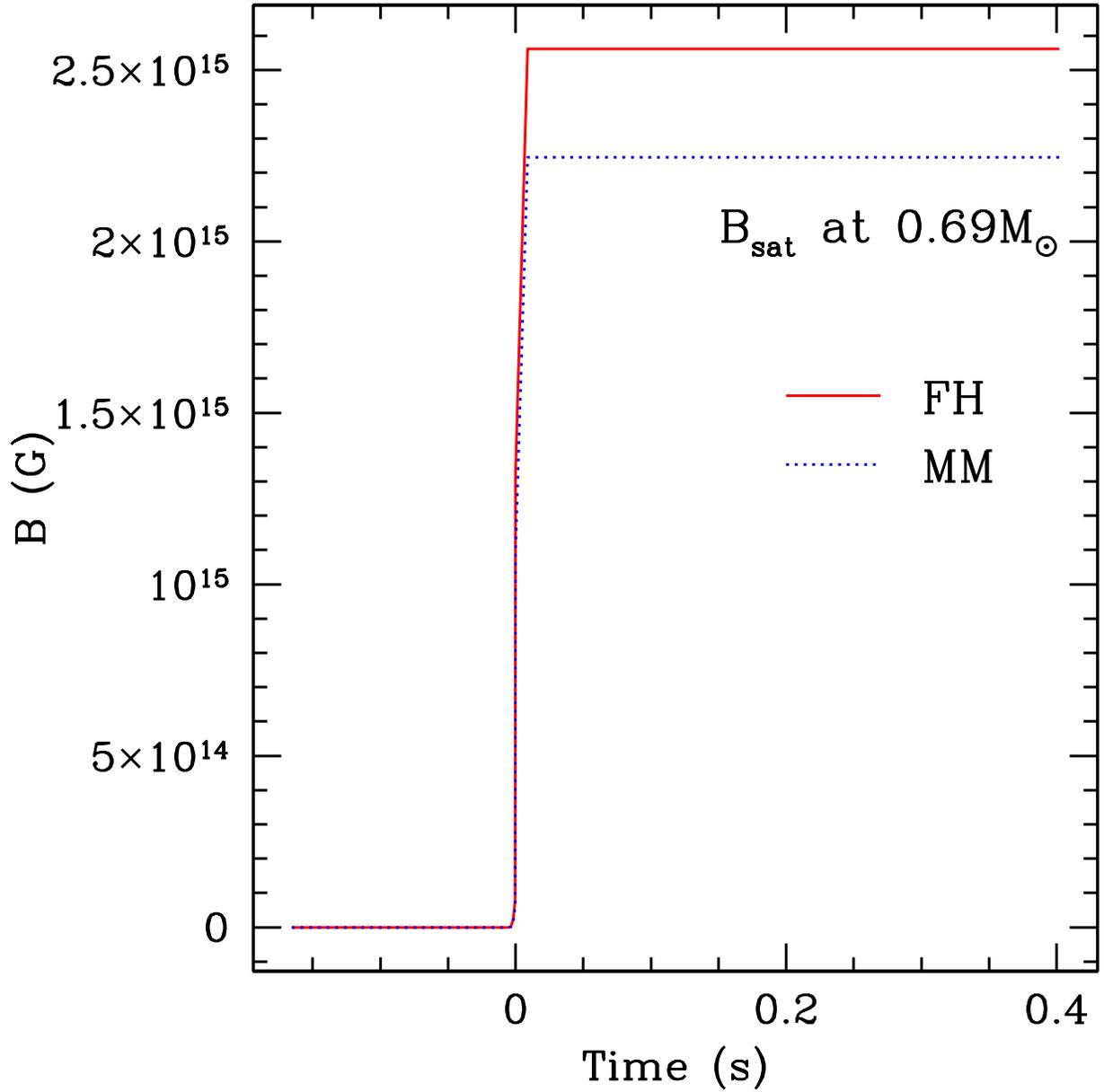}
\figcaption[bvst.eps]{Time evolution of the magnetic field for the
saturation field, $B_{\rm{sat}}$, for the MM and FH profiles at 0.69
$M_{\sun}$ which contains the initial homologous core and the later 
hydrostatic PNS core.  
The field saturates about 9 ms after bounce.\label{bvst}}
\end{figure}

The time evolution of the field with saturation limit $B_{\rm{sat}}$ for
the MM and FH profiles at 0.69 $M_{\sun}$ is plotted in Fig. \ref{bvst}.  
The initial solid body rotation case (\S 3.2) shows similar behavior. This
mass shell contains the initial homologous core at bounce and later the
hydrostatic core of the PNS. The radius of this mass shell is not constant.  
For this mass shell, the magnetic field is amplified shortly before bounce,
and the amplitude quickly jumps to a few times $10^{15}$ G.  The saturation
field is reached in about 9 ms.  In general, the quantity $\Delta B$ in eq.
(\ref{delb}) becomes smaller with time and becomes zero when the field
reaches the saturation limit. In practice, if the shear drops so that the
saturation field decreases, the field can decline as it approaches the
saturation limit.  There is no field dissipation in our treatment, but we
do see the field peak then relax to a lower level in some portions of the
PNS.

For the cases with initial differential rotation, the peak values of the
magnetic field at the end of our calculation at 387 ms after bounce are
$B_{\rm{sat}} = 3.5 \times 10^{16}$ G and $B_{\rm{max,en}} = 4.5 \times
10^{15}$ G for the FH profile, and $B_{\rm{sat}} = 3.3 \times 10^{16}$ G
and $B_{\rm{max,en}} = 3.9 \times 10^{15}$ G for the MM profile.  The
amplitude of the magnetic field is remarkably high and above the QED limit
($B_{\rm{Q}} = 4.4 \times 10^{13}$ G), but remains less than equipartition.  
For the case of $B_{\rm{sat}}$, $P_{\rm{mag}}$ is $\sim$ 10\% of
$P_{\rm{gas}}$, and magnetic buoyancy may limit growth of the magnetic
field \citep{wheel02}.

\subsection{Initial Solid Body Rotation Case}

Even for an initial solid body rotating iron core, collapse generates a
differential rotation profile similar to that of the FH and MM profiles
(Figs. \ref{difrot}, \ref{srot}).  The value of $\Omega(r)$ depends on
$\Omega_{\rm{0\_c}}$: for the solid body case, since we limit our study to
$\Omega_{\rm{0\_c}} \leq 0.2$ rad s$^{-1}$, the angular velocity of the
PNS remains only of order 100 rad s$^{-1}$.  For the solid body case, the
ratio $\Omega/\Omega_{\rm{kep}}$ is expected to peak in the outer regions,
since for a given $\Omega_{\rm{0\_c}}$ the rotational velocity is much
larger at larger radius for solid body rotation than for differential
rotation.  Since in this outer region all three profiles have similar
specific angular momentum, the ratio $\Omega/\Omega_{\rm{kep}}$ is also
similar among these three for our choices of $\Omega_{\rm{0\_c}}$.

Despite the smaller initial angular velocity in the initial solid rotation
case, the magnetic field is amplified to a large value for the
$B_{\rm{sat}}$ case (Fig. \ref{solidb}).  At 387 ms after bounce, the peak
value of $B_{\rm{sat}}$ is $9.0 \times 10^{15}$ G and $2.3 \times 10^{14}$ 
G for $B_{\rm{max,en}}$.


The ratio $\Omega/\Omega_{\rm{kep}}$ peaks just inside of the stalled
shock, but the resulting magnetic field is not maximized at this location.  
The maximum magnetic field is obtained at the boundary of the PNS where
the shear is the strongest.  This is a good illustration of how the shear,
rather than angular momentum, powers the MRI .  Since the magnitude of the
shear is smaller than in the cases of the MM and FH profiles with
$\Omega_{\rm{0\_c}} = 1.0$ rad s$^{-1}$ (Fig.  \ref{shear}), the magnetic
field strength is also smaller.

\subsection{MHD Outflow Power}

We expect that the magnetic field generated by the MRI will power MHD
bi-polar outflow.  Employing the characteristic power of a Blandford \&
Payne type MHD outflow (eq. (\ref{lmhd})), the outflow luminosity
$L_{\rm{MHD}}$ and the dimensionless parameter $\nu$ are calculated for
the three initial rotational profiles, MM, FH, and solid body. 
The profiles of this MHD luminosity mimic those of the magnetic field.  
The parameter $\nu$ is always less than 0.11 for all cases, verifying our 
assumption that the resulting flows would be of the non-relativistic 
Blandford-Payne type.  Fig. \ref{lum} gives the MHD luminosity and the 
parameter $\nu$ at different times for the MM and FH profiles using 
the saturation prescription $B_{\rm{max,en}}$.  For the cases with initial 
differential profiles, the peaks of the MHD luminosity are at the boundary 
of the PNS which coincides with peaks in $\nu$ for the MM and FH profiles.  
The parameter $\nu$ peaks inside the stalled shock for the initial solid 
rotation case due to the relatively higher MHD luminosity and lower critical 
luminosity in that region.  

Our calculations are limited to sub--Keplerian rotation and
sub--equipartition fields, and yet they potentially produce significant MHD
luminosity:  for the saturation field $B_{\rm{sat}}$, the maximum values
387 ms after bounce are 6.6 $\times 10^{53}$ erg s$^{-1}$ for FH, 4.2
$\times 10^{53}$ erg s$^{-1}$ for MM, and 6.5 $\times 10^{51}$ erg s$^{-1}$
for the initial solid rotation.  For the saturation field
$B_{\rm{max,en}}$, the peak values of MHD luminosity 387 ms after bounce
are 1.4 $\times 10^{52}$ erg s$^{-1}$ for FH, 9.0 $\times 10^{52}$ erg
s$^{-1}$ for MM, and 2.7 $\times 10^{48}$ erg s$^{-1}$ for initial solid
rotation.  The investigation of how the MHD luminosity can be turned into a
bi-polar flow is left for future work, although we outline some
possibilities in the discussion below.

\section{Discussion and Conclusions}

No one would doubt that the progenitors of core collapse supernovae 
rotate and possess some magnetic field.  Pulsars as remnants of core
collapse are manifestly rotating and magnetized.  The question has always
been whether rotation and magnetic fields would be incidental
perturbations or a critical factor in understanding the explosion.  We
have shown here that with plausible rotation from contemporary stellar
evolution calculations and any finite seed field with a component parallel
to the rotation axis, the magnetorotational instability can lead to the
rapid exponential growth of the magnetic field to substantial values on
times of a fraction of a second, comparable to the core collapse time.  
Even a relatively modest value of initial rotation gives a very 
rapidly rotating PNS and hence strong differential rotation with 
respect to the infalling matter.

This result seems reasonably robust.  The reason is that the instability
condition for the MRI is basically only that the gradient in angular
velocity be negative.  This condition is broadly satisfied in core
collapse environments.  Rotation can weaken supernova explosions without
magnetic field \citep{FH00}; on the other hand, rotational energy can be
converted to magnetic energy that can power MHD bi-polar flow that may promote
supernova explosions.  {\it The implication is that rotation and magnetic
fields cannot be ignored in the core collapse context.}

Even artificially limiting the post-collapse rotation to sub--Keplerian
values as done here, we find fields in excess of $10^{15}$ G near the
boundary of the neutron star.  While this field strength is
sub--equipartition, the implied MHD luminosities we have derived are of
order $10^{52}$ erg s$^{-1}$.  This is a substantial luminosity and could,
alone, power a supernova explosion if sustained for a sufficiently long
time, a fraction of a second.  As pointed out by Wheeler, Meier \& Wilson
(2002), the fields do not have to be comparable to equipartition to be
important because they can catalyze the conversion of the large reservoir
of rotational energy into buoyant, bi-polar MHD flow.  Higher rates of
initial rotation that are within the bounds of the evolutionary
calculations could lead to even larger post-collapse rotation and even
larger magnetic fields.  If the initial rotation of the iron core proves
to be substantially lower than we have explored here, then the MRI would
be of little consequence to the explosion.  The MHD luminosities derived
here are comparable to the typical neutrino luminosities derived from core
collapse, $\sim 10^{52}$ erg~s$^{-1}$. One important difference is that
the matter beyond the PNS is increasingly transparent to this neutrino
luminosity, whereas the MHD power is deposited locally in the plasma.  
Another difference is that the neutrino luminosity is basically radial so
it resists the inward fall of the collapse, the very source of the
neutrino luminosity itself.  In contrast, hoop stresses associated with
the magnetic field (see below) will tend to pull inward and force matter
selectively up the rotation axis.

We note that for complete self-consistency, one should apply the MRI to
the evolution of rotating stars where even a weak field renders the
H{\o}iland dynamical stability criterion ``all but useless" in the words
of Balbus \& Hawley (1998).  Some steps to explore magnetic viscosity in
stellar evolution have been taken. \citet{spruit98} have invoked a
magnetic viscosity that efficiently couples the inner core to the outer
expanding giant envelope until the core contraction is more rapid than its
rotation. They concluded that if the decoupling is as late as the phase of
carbon depletion, the iron core might be rotating with an extremely low
angular velocity, $\sim 10^{-4}$ rad s$^{-1}$, yielding an initial PNS
rotation period of 100 s or an angular velocity of $\sim$ 0.06 rad
s$^{-1}$.  This condition requires that the field be of order the rotation
energy of the core, but if this condition holds then field amplification
by the MRI in the collapse would be negligible. On the other hand, recent
calculations by \citet{HW02} based on a prescription for magnetic
viscosity by \citet{spruit02} yield much more rapidly rotating iron cores.
\citet{HW02} find PNS rotation rates of 4 to 8 ms, consistent with the
values we have explored here.  Clearly, much more must be done to
understand the magnetorotational evolution of supernova progenitors.

As expected, the shear and hence the saturation fields are often highest
at the boundary of the PNS where strong MHD activity is anticipated.  
Unexpectedly, with our fiducial saturation field, the field can also be
large within the standing shock compared to the local pressure since the
shock compression there naturally leads to shear in a rotating environment
and since gas pressure is relatively low.  The strength of this secondary
peak is about the same for the initial solid body profile with
$\Omega_{\rm{0\_c}} = 0.2$ rad s$^{-1}$ and for the MM and FH profiles
with $\Omega_{\rm{0\_c}} = 1.0$ rad s$^{-1}$.  We have had to limit
$\Omega_{\rm{0\_c}}$ to smaller values in the case of initially constant
angular velocity in order to not violate the sub--Keplerian condition at
larger radii after collapse.  The maximum magnetic fields achieved are
generally about the same strength within a factor of 10 for all three
initial angular velocity prescriptions we have explored, although the
value of $\Omega_{\rm{0\_c}}$ was chosen to be substantially smaller in
the initially solid rotation case compared to those where there is a
gradient in $\Omega$.  

The configuration of the magnetic field in a precollapse iron core is not
well understood.  In this calculation we have assumed there exists a seed
vertical field to calculate the growth of the field due to the MRI;  
however, the MRI can amplify other components of the magnetic field.  The
final configuration of the magnetic field after collapse may be less
uncertain since the system has a strongly preferred direction due to
rotation.  In the context of accretion disks, \citet{HGB96} show that
their fiducial run with initial random magnetic field configuration
results in 9\% radial, 88 \% toroidal, and 3 \% vertical components.  
Most of the shear is in the radial direction, so the radial component is
greatly amplified by the MRI and turned into toroidal field due to
differential rotation \citep{BH98}.  The dominant component is most likely
to be the toroidal field.

Another uncertainty is the rotational profile.  Although we have formally
considered the MRI in a geometry with arbitrary pressure gradients, in
practice our assumption of conservation of angular momentum in spherical
shells effectively restricts our analysis of the resulting magnetic fields
to the equatorial plane.  It is not clear what profile to use in the PNS,
since, we note, even the rotational profile of the Sun is not well
understood.  A full understanding of the rotational state of a PNS remains
a large challenge.

We have assumed various prescriptions for the saturation field.  All are
variations on the theme that, within factors of order 2$\pi$, the
saturation field will be given by the condition $v_{\rm{A}} \sim r
\Omega$.  Fig. \ref{vel} compares the velocities of sound ($c_{\rm{s}}$),
Keplerian rotation ($v_{\rm{kep}}$) and model rotation ($r\Omega$), and
the Alfv\'{e}n ($v_{\rm{A}}$) velocity corresponding to $B_{\rm{sat}}$ for
the solid, MM and FH profiles we adopted.  For the sub--Keplerian
rotational velocity profiles we employed, the magnetic field saturates at
$v_{\rm{A}} \sim r\Omega$.  For the case of $B_{\rm{max,en}}$, $v_{\rm{A}}
< r\Omega$ since $B_{\rm{max,en}} < B_{\rm{sat}}$ for the MM and FH
profiles, and $B_{\rm{max,en}}$ is especially small for the initial solid
rotation profile.  In all cases, the saturation field is
sub--equipartition with Alfv\'{e}n velocity less than Keplerian.  In the
numerical disk simulations, about 20 rotations are required to reach
saturation.  The region of maximum shear in these calculations, around 15
km, typically has an angular velocity of 500 rad s$^{-1}$ or a period of
about 0.013 s.  That means that by the end of the current calculations at
0.387 s, there have been about 30 rotations. Although the prescriptions
for the growth and saturation fields we use here are heuristic, this
aspect of our results is certainly commensurate with the numerical
simulations of the MRI.
  
The issues of the saturation field and the nature of astrophysical dynamos
are still vigorously explored.  There are concerns that even weak fields
may limit turbulent cascade at the smallest scales and hence suppress
dynamo processes.  Vainshtein, Parker \& Rosner (1993) point out that this
may be avoided by the formation of Coriolis-twisted loops that develop in
shearing strong toroidal fields.  Papaloizou and Szuszkiewicz (1992)
presented a global stability analysis for conditions where the angular
velocity is constant along magnetic lines of force that generally agreed
with the local MRI stability analysis of Balbus \& Hawley (1998) for weak
fields.  For strong fields, however, their analysis suggested that fields
corresponding to the saturation condition $v_{\rm{A}} \sim \sqrt{R \Omega
c_{\rm{s}}}$ would be unstable.  For the sub--Keplerian conditions we
consider here, this saturation criterion would yield a ratio of magnetic
to total pressure that is larger by a factor of $c_{\rm{s}}/r \Omega$ than
the criteria we used (see Fig.  \ref{vel}). \citet{VC01} conclude that the
MRI has the required properties for a dynamo, anisotropic turbulence in a
shearing flow, to generate both disordered and ordered fields of large
strength.  The saturation limits we have adopted here are consistent with
those found in numerical calculations of the MRI saturation, but this
topic clearly deserves more study.

We have made one major assumption that must be checked more carefully in
future work.  That is that we can apply the MRI linear stability analysis
derived for no background radial flow to a dynamic situation where radial
inflow and convective motions are rapid.  The character of the MRI has
never been investigated in this regime with finite background flow.  One
rationale for our assumption is that at a given radius one can transfer to
a co-moving frame where the local radial velocity is zero.  This is
formally correct, but since we then make assumptions about modes with
wavelength comparable to the size of the system, this formality may be
inapplicable to the problem we want to solve, the nature of the saturation
fields.  An ex-post factor justification for our results is that after
collapse when the PNS settles nearly into hydrostatic equilibrium, the
radial motions are, in fact, small and the traditional MRI is probably
valid as illustrated in Fig. \ref{vrad}. The action of the MRI in the
vicinity of the standing shock and the fields we derive there may,
however, be questionable on this basis.  This topic deserves more study as
a general point of physics, not merely for the application to core
collapse. We also note that the MRI could be altered in the context of
large radiation pressure.  The radiation pressure is not a major component
of the total pressure in the current calculations, so we do not believe
this to be a factor here, but this is worth considering in principle.
  
We have also neglected several feed-back processes. Among these are the
build up of magnetic pressure and hoop stresses, the viscous coupling of
shells that will tend to suppress the differential rotation, the effects
of neutrino viscosity on turbulence, the effects of the magnetic field on
neutrino transport, and the effects of centrifugal forces.  We will
describe some of these issues briefly.


Both the magnetic pressure and the magnetic viscosity are small for the
sub--Keplerian conditions explored here.  For most cases $\beta^{-1}$ is
less than 0.1 for the conditions we have assumed, (the $B_{\rm{sat}}$ case
for the FH profile with $\Omega_{0\_\rm{c}} = 1.0 s^{-1}$ pushes this
limit), so the direct dynamical effect of the magnetic field is expected to
be small.  The viscous time scale is $\tau_{\rm{vis}} \sim (\alpha
\Omega)^{-1}(\rm{r/h})^2$, where $\alpha$ is the viscosity parameter and h
is the vertical scale height, with h $\sim$ r for our case.  For a
magnetically-dominated viscosity,
\begin{equation}
\alpha \sim \frac{B_{\rm{r}} B_{\phi}}{4 \pi P} =
\left(\frac{B_{\rm{r}}}{B_{\phi}}\right) \frac{B_{\phi}^2}{4 \pi P} \sim 2
\left(\frac{B_{\rm{r}}}{B_{\phi}}\right) \beta^{-1}.
\end{equation}
With this expression for $\alpha$, the viscous
time becomes:
\begin{equation}
\tau_{\rm{vis}} \sim \frac{1}{2}\left(\frac{B_{\phi}}{B_{\rm{r}}}\right)
\left(\frac{1}{\beta^{-1}\Omega}\right) \gg \Omega^{-1},
\end{equation}
where the final inequality follows from $B_{\phi} > B_{\rm{r}}$ and
$\beta^{-1} < 1$.  This prescription for viscosity is
reasonable in the absence of convection.  In the portions of the structure
that are convective, the viscosity could be enhanced significantly.

In addition to dynamic effects, we note that the fields generated here are
well above the QED limit ($B_{\rm{Q}} = 4.4 \times 10^{13}$ G).  In this
exotic regime, such a strong field has radiative and thermodynamic effects
\citep{duncan00}, although it is not clear that these have profound
effects on the dynamics in core collapse supernovae.  

We have not discussed the role of neutrinos here, although the processes
of neutrino loss and de-leptonization are included in our calculation of
the cooling PNS.  It is possible that the neutrino flux affects the
magnetic buoyancy (Thompson \& Murray 2002) and that the magnetic fields
affect the neutrino emissivity \citep{TD96} and interactions with the
plasma (Laming 1999).  The time scale for shear viscosity due to neutrino
diffusion is much longer than the times of interest here, although
magnetic fields and turbulence can make it shorter \citep{gou98}.  The MRI
provides magnetic field and turbulence, so this issue deserves further
study.  In addition to affecting the shear, the neutrino viscosity might
also affect the turbulence needed to make the MRI work.  Using the
expression for the neutrino viscosity at sub-nuclear densities from
\citet{gou98},
\begin{equation}
\label{nuvisc}
\eta_{\nu} \sim 2\times10^{23}\left(\frac{T}{10~{\rm
MeV}}\right)^2\rho_{13}^{-1} {\rm g~cm^{-1}~s^{-1}},
\end{equation}
gives a Reynolds number 
\begin{equation}
\label{reynolds}
Re \sim \frac{v R \rho}{\eta_{\nu}} \sim \frac{\Omega R^2 \rho}{\eta_{\nu}} \sim 10^5 
\end{equation}
near the PNS boundary at $R \sim 10^6$ cm with $\Omega \sim 1000$ rad
s$^{-1}$.  This could decrease to Re $\sim 100$ near the standing shock
where the density is lower.  The former is probably large enough to
sustain turbulence in the strong shear, but the latter might not be.

When viscosity dominates the dissipation in a collisional plasma, the growth
condition that the field growth time be less than the dissipation time yields
a constraint on the magnetic field \citep{BH98}:
\begin{equation}
B \gg \left(\frac{15 \pi}{8} \nu \rho \Omega\right)^{1/2} = 5\times10^{7}~\rm{G}~
\Omega_3^{1/2}\left(\frac{T}{10~{\rm MeV}}\right)^{5/4} 
\end{equation}
This equation implies that B must exceed $\sim 10^8$ G near the boundary
of the PNS where T $\sim 10$ MeV and smaller values in cooler portions at
larger radii.  Even compression of moderate fields in the iron core should
exceed this threshold. We note that if one were to use the expression for
the neutrino viscosity from eq. (\ref{nuvisc}), the corresponding limit on
the field would exceed $10^{13}$ G for similar parameters.  This limit may
not be relevant, however, since in order to use the fluid equations,
rather than a kinetic theory, to describe the instability and its damping,
the physical lengthscale associated with the viscosity, the mean free path
of the dissipative particles, must be substantially less than the damping
scale of the turbulence.  This is not likely to be the case for neutrinos
in the vicinity of the region of maximum shear in this problem.  The
damping length where Re $\sim$ 1 would be about $10^{-5}$ of $10^6$ cm or
about 10 cm from eq. (\ref{reynolds}), whereas the mean free path of the
neutrinos (neglecting blocking and other complications) is about $10^4$ cm
at the density of $\sim 10^{13}$ {g~cm$^{-3}$} and temperature of about 10
MeV that characterize the region of maximum shear \citep{arnett96}.  
There may be regions deep in the neutron star where the neutrino mean free
path is short enough that this becomes an issue, but we do not think that
the neutrinos can affect the eddy turnover and instability of the MRI in
regions that are significant for maximum field growth.
    
An obvious imperative is to now understand the behavior of the strong
magnetic fields we believe are likely to be attendant to any core collapse
situation.  The fields will generate strong pressure tensor anisotropies
that can lead to dynamic response even when the magnetic pressure is small
compared to the isotropic ambient gas pressure. As argued in Wheeler,
Meier \& Wilson (2002), a dominant toroidal component is a natural
condition to form a collimated magnetorotational wind, and hence polar
flow.  A first example of driving a polar flow with the MRI is given by
\citet{HB02}.  A key ingredient to force flow up the axis and to collimate
it is the hoop stress from the resulting field.  We have examined the
acceleration implied by the hoop stresses of the fields we have derived
here, $a_{\rm{hoop}} = B_{\phi}^2/4\pi \rho r$.  We find that the hoop
stresses corresponding to the peak saturation fields can be competitive
with, and even exceed, the net acceleration of the pressure gradient and
gravity.  The large scale toroidal field is thus likely to affect the
dynamics by accelerating matter inward along cylindrical radii.  The flow,
thus compressed, is likely to be channeled up the rotation axes to begin
the bi-polar flow that will be further accelerated by hoop and torsional
stresses from the field, the ``spring and fling" outlined in Wheeler et
al. (2002). The MRI is expected to yield a combination of large scale and
small scale magnetic fields.  Even the small scale fields in the turbulent
magnetized medium may act like a viscoelastic fluid that would tend to
drive circulation in along the equator and up the rotation axis
\citep{williams02} where the same small scale fields could collimate and
stabilize the flow even in the absence of large scale toroidal fields
\citep{li02}. On the other hand, small scale fields can result in
dissipation by reconnection, an issue that we have not treated here,
relying implicitly on the numerical simulations that show a growth in the
field that is not eliminated by such effects.

The dynamics of these jets may depart substantially from pure
hydrodynamical jets, since they will tend to preserve the flux in the
Poynting flow and reconnection can accelerate the matter (see Spruit,
Daigne \& Drenkhahn 2001 and references therein). The poloidal component
of the field can be another contributor to plasma waves (Wheeler et al.
2000).

For a complete understanding of the physics in a core collapse supernova
explosion, a combination of neutrino--induced and jet--induced explosion
may be required.  Understanding the myriad implications of this statement
will be a rich exploration.

\acknowledgments We thank Steve Balbus, Roger Blandford, Stirling Colgate,
Peter H\"oflich, Julian Krolik, Pawan Kumar, John Scalo, Jim Stone, Chris
Thompson, Ethan Vishniac, Peter Williams and Ellen Zweibel for valuable
discussions.  We also thank the anonymous referee for several valuable
comments.  JCW is grateful to the hospitality of the Aspen Center for
Physics where this work germinated.  This work was supported in part by
NASA Grant NAG59302 and NSF Grant AST-0098644.






\clearpage

\clearpage

\clearpage

\clearpage

\clearpage

\clearpage

\clearpage

\clearpage

\clearpage

\clearpage

\clearpage

\begin{figure}[htp]
\centering
\includegraphics[totalheight=0.8\textheight]{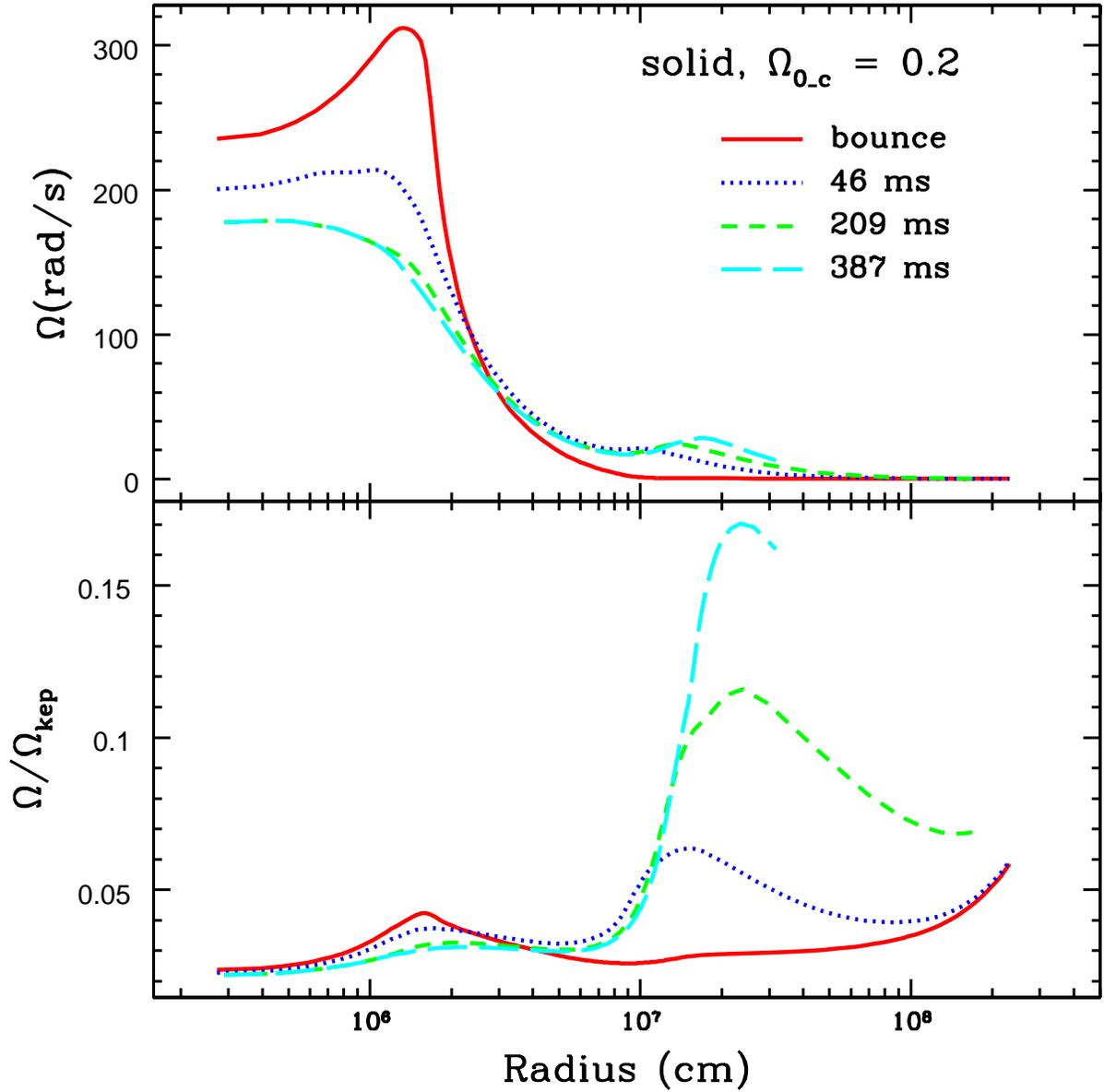}
\figcaption[solidomega.eps]{Rotational profile and
$\Omega/\Omega_{\rm{kep}}$ for the initial solid body case with
$\Omega_{\rm{0\_c}} = $0.2 rad s$^{-1}$.  The differential rotation
profile is similar to those for the initial differential rotation
cases.\label{srot}}
\end{figure}

\clearpage

\begin{figure}[htp]
\centering  
\includegraphics[totalheight=0.8\textheight]{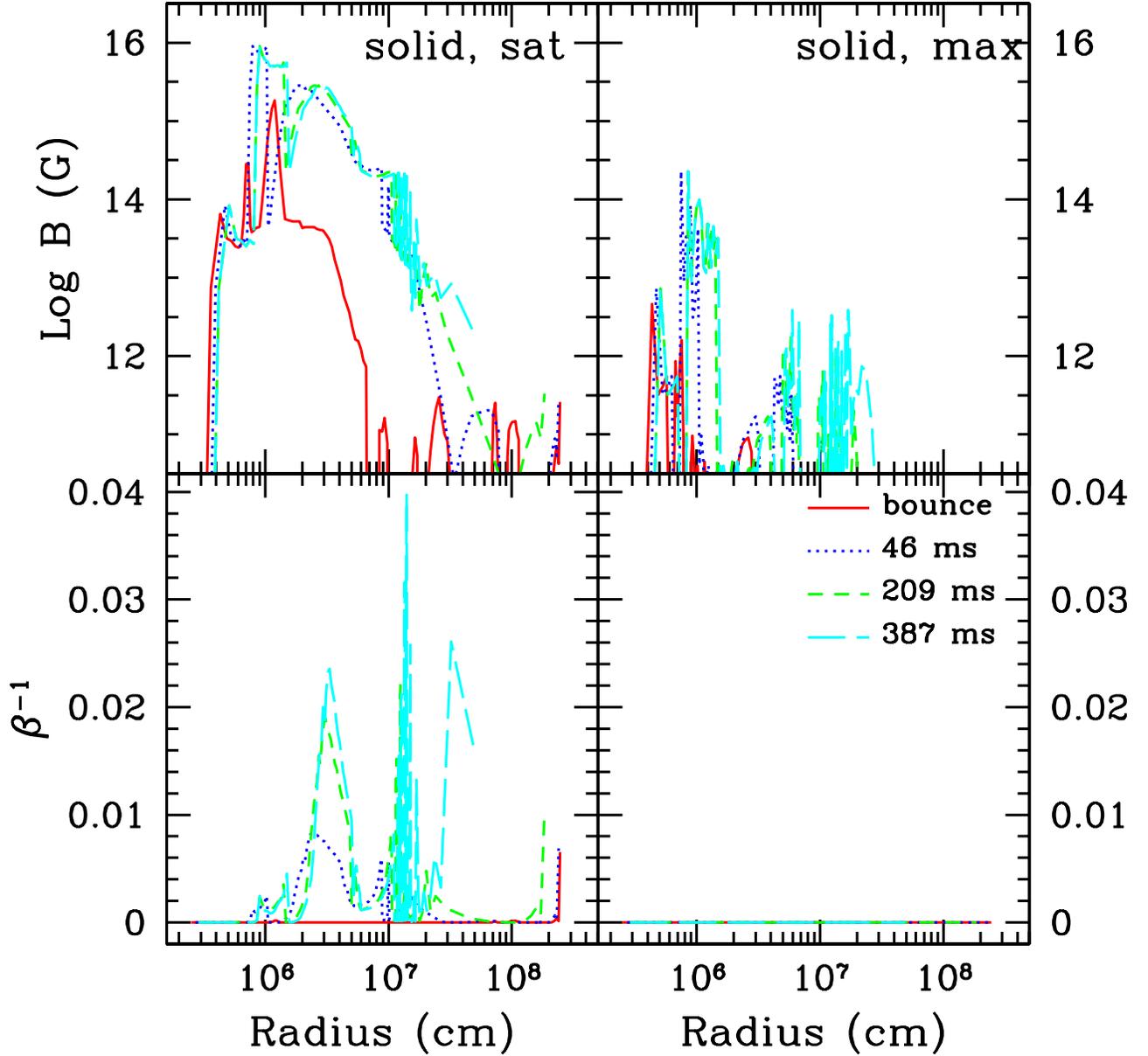}
\figcaption[solidb.eps]{The magnetic field distribution and $\beta^{-1}$
for the fiducial saturation field $B_{\rm{sat}}$ and the saturation
field that corresponds to the maximum growing mode of the MRI
$B_{\rm{max,en}}$ for the initial solid rotation case.  Despite the smaller
$\Omega_{\rm{0\_c}}$, the maximum value for the $B_{\rm{sat}}$ case is
$B \sim 10^{15}$ G and $B \sim 10^{14}$ for the $B_{\rm{max,en}}$ case 387 
ms after bounce.  The magnetic field for the $B_{\rm{max,en}}$ case is 
less significantly amplified because of the small angular velocity and small 
shear. \label{solidb}} \end{figure}

\clearpage

\begin{figure}[htp]
\centering
\includegraphics[totalheight=0.8\textheight]{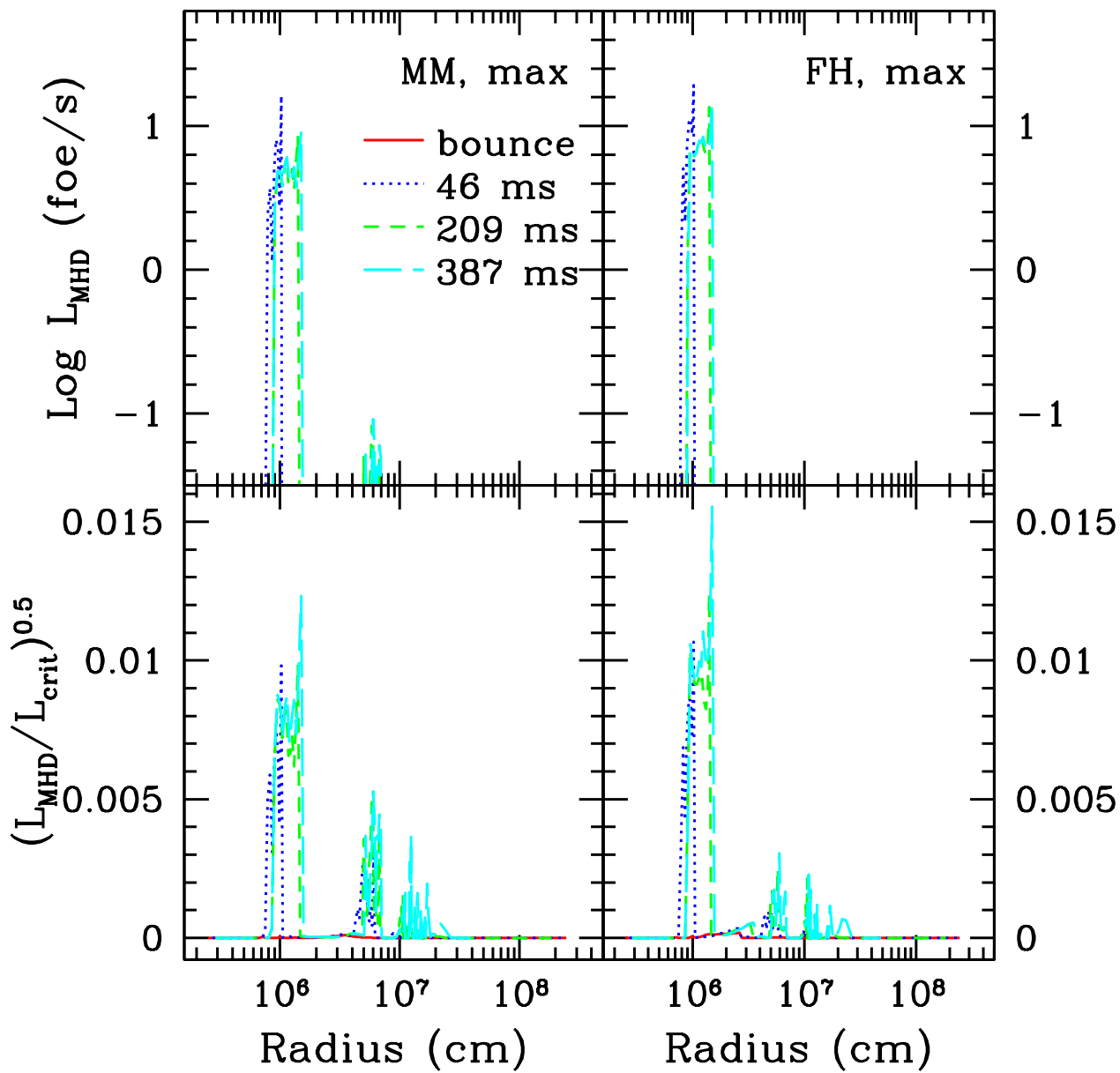}
\figcaption[lum.eps]{MHD luminosity (1 foe = $10^{51}$ erg)  
for Blandford \& Payne type outflow and $\nu 
\equiv (L_{\rm{MHD}}/L_{\rm{crit}})^{0.5}$ for the saturation field 
$B_{\rm{max,en}}$ for the differential rotation profiles (MM and FH).  
The maximum value of $L_{\rm{MHD}}$ is $\sim 10^{52}$ erg s$^{-1}$
for both cases 387 ms after bounce.\label{lum}}
\end{figure}

\clearpage

\begin{figure}[htp]
\centering
\includegraphics[totalheight=0.8\textheight]{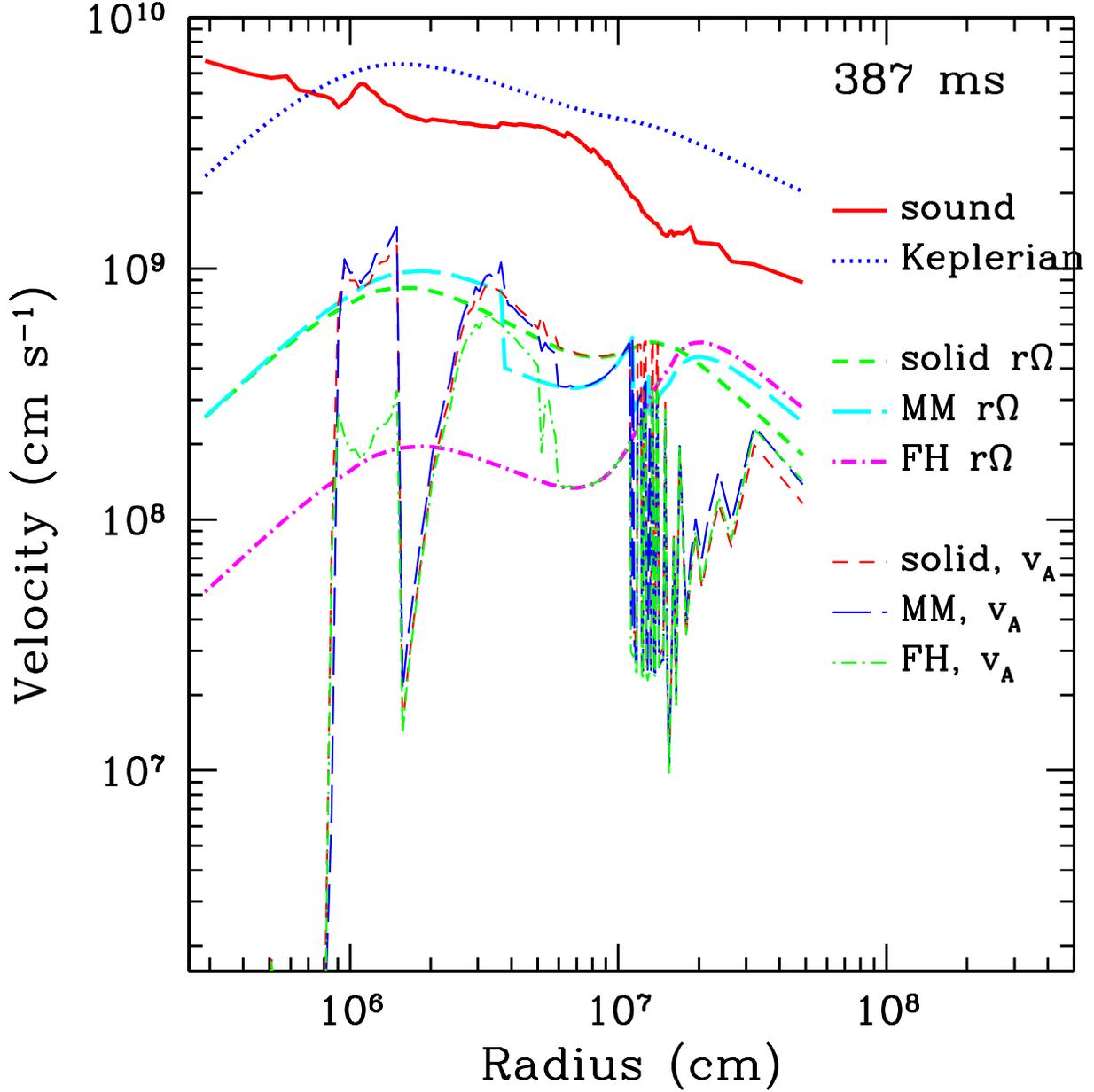}
\figcaption[vel.eps]{Comparison of velocities of sound ($c_{\rm{s}}$),
Keplerian rotation ($v_{\rm{kep}}$), model rotation ($r\Omega$), and
the Alfv\'{e}n ($v_{\rm{A}}$) velocity for the saturation field $B_{sat}$
for the three rotational profiles.  At 387 ms after bounce, the magnetic
field saturates at $v_{\rm{A}} \sim r\Omega$, which is sub--Keplerian and
represents a sub--equipartition field.\label{vel}}
\end{figure}

\clearpage

\begin{figure}[htp]
\centering
\includegraphics[totalheight=0.8\textheight]{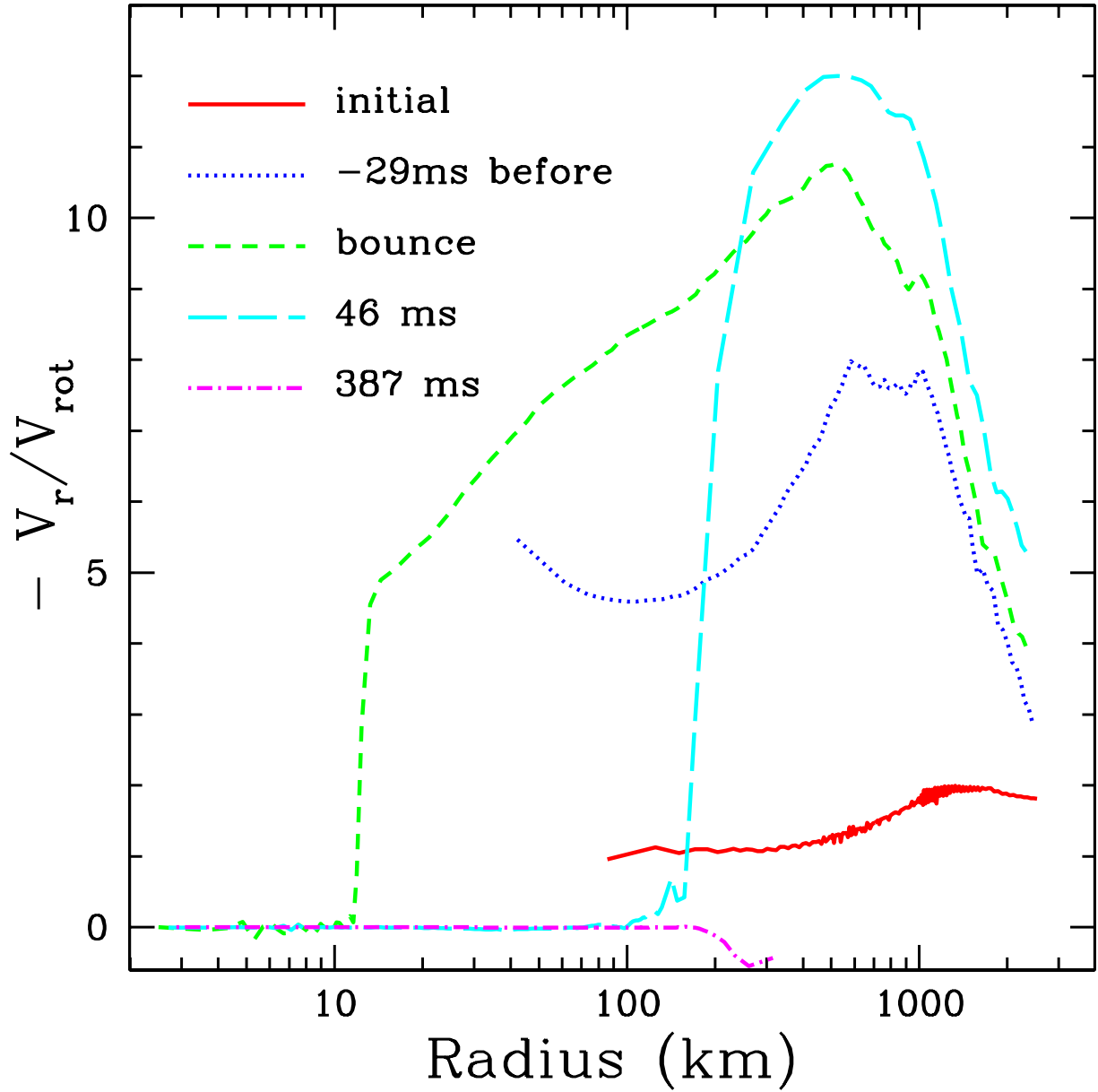}
\figcaption[radialv.eps]{The ratio of the radial velocity to the deduced
rotational velocity for the MM initial profile is shown as a function of
time.  For times after $\sim 50$ ms, the radial velocity is low out to
$\sim 100$ km and the treatment of the MRI instability neglecting background 
velocities is justified. \label{vrad}}
\end{figure}


\end{document}